%% file: main.tex
\documentclass[twocolumn]{aastex701}
\usepackage{amsmath,amssymb}
\usepackage{xcolor}
\usepackage{enumitem}
\usepackage[acronym]{glossaries}
\usepackage[nameinlink]{cleveref}
\usepackage{graphicx} % figures
\usepackage{placeins} % figure floatbarrier

\usepackage{tikz}
\usetikzlibrary{arrows.meta, calc}

\usepackage{preamble/preamble-starkman}

\crefname{equation}{Eq.}{Eqs.}
\Crefname{equation}{Eq.}{Eqs.}
\crefname{figure}{Figure}{Figures}
\Crefname{figure}{Figure}{Figures}

\newacronym{des}{DES}{Dark Energy Survey}
\newacronym{sgl}{SGL}{SGL}
\newacronym{NFW}{NFW}{Navarro--Frenk--White}
\newacronym{LMC}{LMC}{Large Magellanic Cloud}
\newacronym{Sgr}{Sgr}{Sagittarius}
\newacronym{Rubin}{Rubin}{Vera C. Rubin Observatory}
\newacronym{Roman}{\textsl{Roman}}{Nancy Grace Roman Space Telescope}
\newacronym{ARRAKIHS}{\textsl{ARRAKIHS}}{Analysis of Resolved Remnants of Accreted galaxies as a Key Instrument for Halo Surveys}
\newcommand{\rubinlsst}{\gls{Rubin}/\gls{LSST}}

\defcitealias{Nibauer+:2023}{N23}

\makeglossaries
\glsdisablehyper

\newcommand{\Xvec}{\mbf{X}}  % 3D position vector
\newcommand{\Zhat}{\hat{Z}}

\newcommand{\zhat}{\hat{\mbf{e}}_z} % unit vector in z direction
\newcommand{\xvec}{\mbf{x}}  % 3D position vector
\newcommand{\vel}{\mbf{v}}  % 3D velocity vector
\newcommand{\avec}{\mbf{a}}  % 3D acceleration vector
\newcommand{\that}{\hat{\mbf{t}}}  % 3D tangent unit vector
\newcommand{\nhat}{\hat{\mbf{n}}}  % 3D Frenet-Serret normal unit vector

\newcommand{\xcom}{\proj{\bar{\xvec}}}  % reference position vector
\newcommand{\Astr}{\mathcal{A}_{\mathrm{str}}}

\newcommand{\sarc}{s}  % 3D arc length scalar
\newcommand{\speed}{v}  % 3D speed scalar
\newcommand{\curvature}{\kappa}  % 3D curvature scalar
\newcommand{\torsion}{\tau}  % 3D torsion scalar

\newcommand{\proj}[1]{{#1}_{p}}  % projected quantity
\newcommand{\projperp}[1]{{#1}_{p \perp}}  % projected quantity
\newcommand{\project}[1]{\mrm{pr}\!\left({#1}\right)}  % projection operator

\newcommand{\xsky}{\proj{\xvec}}  % planar position vector (x,y)
\newcommand{\vsky}{\proj{\vel}}  % projected velocity vector
\newcommand{\asky}{\proj{\avec}}  % projected acceleration vector
\newcommand{\tphat}{\proj{\that}}  % projected tangent unit vector
\newcommand{\nphat}{\proj{\nhat}}  % projected normal unit vector

\newcommand{\xskyperp}{\projperp{\xvec}}

\newcommand{\ssky}{\proj{\sarc}} % projected arc length
\newcommand{\curvaturesky}{\proj{\curvature}}  % projected curvature
\newcommand{\speedsky}{\proj{\speed}}

\newcommand{\lamM}{\ensuremath{\lambda_{\mcal{M}}}}   % rate under model M
\newcommand{\lamC}{\ensuremath{\lambda_{\mrm{C}}}}    % CDM rate
\newcommand{\lamSI}{\ensuremath{\lambda_{\mrm{SI}}}}  % SIDM rate
\newcommand{\lamObs}{\ensuremath{\hat{\lambda}_{\mrm{obs}}}}  % observed rate K/N
\newcommand{\lamdadet}[1]{\ensuremath{\lambda_{\rm det}^{({#1})}}}

\newcommand{\script}[1]{}

\newcommand{\lamCval}{\input{output/cosmology/lambda_arc_cdm.tex}\%}
\newcommand{\lamSIval}{\input{output/cosmology/lambda_arc_sidm.tex}\%}

\NewDocumentCommand {\vecb} {m m} {
\left({#2}\right)_{\!\text{\fontsize{4}{4}\selectfont #1}}%
}

\NewDocumentCommand {\vecC} {s m} {
\IfBooleanTF{#1}
{ \left( #2 \right) }
{ \vecb{C}{#2} }
}

\registerpackagename{Potamides}{https://potamides.readthedocs.io}

\newcommand{\starkman}[1]{{\textcolor{orange}{(NS: #1)}}}

\makeatletter
\AtBeginDocument{%
  \let\ltx@label\label
  \def\label@in@display@noarg#1{\cref@old@label@in@display{#1}}%
  \def\label@in@display@optarg[#1]#2{\cref@old@label@in@display[#1]{#2}}%
}
\makeatother

\begin{document}

\title{When Streams Curve Away: a Test of Dark Matter from Extragalactic Stellar Stream Populations}
\shorttitle{When Streams Curve Away: Constraining Dark Matter Halo Geometry}

%% Author information
\author[orcid=0000-0003-3954-3291,sname=Starkman,gname=Nathaniel]{Nathaniel Starkman}
\affiliation{Kavli Institute for Astrophysics and Space Research, Massachusetts Institute of Technology, Cambridge, MA 02139, USA}
\affiliation{Case Western Reserve University, 10900 Euclid Ave, Cleveland, OH 44106}
\email[show]{starkman@mit.edu}

\author[/0000-0001-8042-5794,sname=Nibauer,gname=Jacob]{Jacob Nibauer}
\affiliation{Department of Astrophysical Sciences, Princeton University, Princeton, NJ 08544, USA}
\affiliation{Center for Astrophysics $\vert$ Harvard \& Smithsonian, 60 Garden Street, Cambridge, MA 02138, USA}
\email[hide]{jnibauer@cfa.harvard.edu}

\author[orcid=0000-0003-0256-5446,sname=Pearson,gname=Sarah]{Sarah Pearson}
\affiliation{DTU Space, Technical University of Denmark, Elektrovej 327, DK- 2800 Kgs. Lyngby, Denmark}
\affiliation{Niels Bohr Institute, DARK, University of Copenhagen, Jagtvej 155A, 2200 Copenhagen N, Denmark}
\email[hide]{sarah.pearson@nbi.ku.dk}

\author[orcid=0009-0003-4675-3622,sname=Wu,gname=Sirui]{Sirui Wu}
\affiliation{Niels Bohr Institute, DARK, University of Copenhagen, Jagtvej 155A, 2200 Copenhagen N, Denmark}
\email[hide]{sirui.wu@nbi.ku.dk}

\author[orcid=0000-0003-2806-1414,sname=Necib,gname=Lina]{Lina Necib}
\affiliation{Department of Physics, Massachusetts Institute of Technology, Cambridge, MA 02139, USA}
\affiliation{Kavli Institute for Astrophysics and Space Research, Massachusetts Institute of Technology, Cambridge, MA 02139, USA}
\email[hide]{lnecib@mit.edu}

%% Abstract
\begin{abstract}\label{sec:abstract}

    The nature of dark matter sets the shape of galactic halos: %
    collisionless cold dark matter (CDM) predicts triaxial halos, while scattering in self-interacting dark matter (SIDM) rounds them toward sphericity. %
    We show that a large catalog of projected stellar stream tracks can statistically constrain halo shapes, and therefore the particle nature of dark matter, without detailed modeling of any individual stream or gravitational potential. %
    We develop a new geometric diagnostic for stellar streams: if the curvature of any segment along a stream points away from the host galaxy's projected center of mass, then that stream falsifies broad classes of potentials, including spherical ones. %
    We call such a segment convex. %
    The diagnostic needs only a projected stream track and a host-center estimate, requiring no kinematics, distances, progenitor model, or fit to individual host potentials. %
    Halo triaxiality sets the rate at which streams show convex segments, making this rate an observable of the halo-shape distribution and the underlying dark-matter model. %
    We forecast that upcoming survey-scale stream morphology catalogs can discriminate between CDM and SIDM at up to $5\sigma$. %

\end{abstract}

%% Keywords using Unified Astronomy Thesaurus concepts

%%%%%%%%%%%%%%%%%%%%%%%%%%%%%%%%%%%%%%%%%%%%%%%%%%%%%%%%%%%%%%%%%%%%%%%%%%%%%%%%
%% Body of the manuscript
%%%%%%%%%%%%%%%%%%%%%%%%%%%%%%%%%%%%%%%%%%%%%%%%%%%%%%%%%%%%%%%%%%%%%%%%%%%%%%%%

\section[Introduction]{Introduction}\label{sec:intro}

    % Halo shape and the DM model
    In the \gls{CDM} paradigm, halos are triaxial, with a shape distribution that depends on mass and redshift and with principal axes often misaligned with the baryons, e.g.\ the stellar disk where one is present \citep{Bailin+Steinmetz:2005:InternalExternalAlignment,Vera-Ciro+:2011:ShapeDarkMatter,Baptista+:2022:OrientationsDMHalos}. %
    In simulations, galactic halos have minor-to-major axis ratios $s = c/a \sim \numrange{0.5}{0.7}$, with weak dependence on environment \citep{Jing+Suto:2002:TriaxialModelingHalo,Allgood+:2006:ShapeDarkMatter,Giocoli+:2026:AIDATNGProject3D}. %
    Baryons sphericalize halos more in the inner region, within $\sim 0.1\,R_{200} \approx \qtyrange{20}{30}{\kilo\parsec}$ for Milky-Way-mass systems \citep{Chua+:2019:ShapeDarkMatter,Prada+:2019:DarkMatterHalo,Chua+:2022:ImpactGalacticFeedback,Petit+:2023:ShapeAlignmentMass,Vargya+:2022:ShapesMilkyWayMassGalaxies}. %
    Here $R_{200}$ is the radius enclosing a mean density \num{200} times critical density of the universe. %
    Outside the inner regions halos remain more triaxial, with $c/a \sim \numrange{0.6}{0.7}$ and increasing triaxiality at $r \gtrsim \qty{50}{\kilo\parsec}$, retaining information about the initial cosmological collapse \citep{Kazantzidis+:2004:EffectGasCooling,Kazantzidis+:2010:SphericalizationDarkMatter}. %

    % SIDM shape signal
    Alternative dark matter models predict different shapes. %
    In this work we use \Gls{SIDM} as an example, however the methodology generalizes to other models. %
    \Gls{SIDM} particles scatter and thermalize the inner halo, making the halo rounder than it would be in collisionless \gls{CDM} \citep{Dave+:2001:HaloPropertiesCosmological,Peter+:2013:CosmologicalSimulationsSelfInteracting,Kaplinghat+:2014:TyingDarkMatter}. %
    Halo shapes therefore probe the dark sector \citep[for reviews see][]{Tulin+Yu:2018:DarkMatterSelfinteractions,Adhikari+:2025:AstrophysicalTestsDark}. %
    Cosmological simulations with a dark matter self-interaction cross-section per unit mass of $\sigma/m \sim \qty{1}{\square\centi\meter\per\gram}$ predict measurable differences between \gls{SIDM} and \gls{CDM} axis ratios over much of the halo, from galaxies to clusters \citep{Vargya+:2022:ShapesMilkyWayMassGalaxies,Fischer+:2024:CosmologicalIdealizedSimulations,Brinckmann+:2018:StructureAssemblyHistory}. %
    Baryons shrink this difference, so the shape signal is cleanest outside their dominant region \citep{Despali+:2022:ConstrainingSIDMHalo}. %

    % Weak Lensing
    However, measuring those shapes is hard. %
    Weak gravitational lensing provides one way, through the distortion of background galaxies \citep{Bartelmann+Schneider:2001:WeakGravitationalLensing}. %
    For $\sim\qtyrange{e11}{e13}{\Msun}$ galaxies, however, the signal from an individual galaxy is too weak to provide a useful shape constraint, so weak-lensing analyses stack many systems, aligned by their light, to recover an ensemble-averaged shape \citep{vanUitert+:2012:ConstraintsShapesGalaxy,Schrabback+:2021:TighteningWeakLensing}. %
    Stacking baryon-aligned systems can wash out signal as the outer halo orientation is not expected to couple to the baryons \citep{Bett:2012:HaloShapesWeak}; instead we want the ensemble shape information derived system by system. %

    % Streams as halo-shape tracers
    Stellar streams, the tidal debris of disrupted globular clusters and dwarf galaxies, offer that information. %
    They are excellent tracers of the gravitational potential because their stars occupy orbits near to their progenitors' \citep{Johnston:1998:PrescriptionBuildingMilky,Kupper+:2012:MoreStructureTidal}. %
    Streams with large apocenters probe the outskirts of their host halos, where dark matter dominates the potential. %
    These properties make stellar streams a promising test of \gls{SIDM} halo shape, and \gls{SIDM} zoom-in suites now provide galactic halo populations to predict halo shapes from \citep[e.g.\,][]{Nadler+:2025:SIDMConcertoCompilation}. %
    In the Milky Way, 6D astrometry has enabled detailed stream-based constraints on the Galactic potential and its dark matter halo \citep{Law+Majewski:2010:SagittariusDwarfGalaxy,Koposov+:2010:ConstrainingMilkyWay,Bovy+:2016:SHAPEINNERMILKY,Bonaca+Hogg:2018:InformationContentCold,Vasiliev+:2021:TangoThreeSagittarius,Ibata2023ChartingDR3,Bariego-Quintana:2024:TorsionStellarStreams}. %

    % Imaging surveys
    \glsunset{Rubin}\glsunset{Roman}\glsunset{ARRAKIHS} % Mark as used
    Beyond the Local Group, surveys cannot recover full 6D phase-space information for essentially any stream \citep{Martinez-Delgado+:2010:StellarTidalStreams}, and cannot measure the three-dimensional stream torsion that encodes halo asphericity \citep{Bariego-Quintana:2024:TorsionStellarStreams}. %
    Current and upcoming wide-field imaging surveys will not measure these quantities either. %
    The \euclid{} mission \citep{EuclidCollaboration+:2025:EuclidQuickData,Euclid+Walmsley+:2025:Q1}, the Vera C. Rubin Observatory's \gls{LSST} \citep{Ivezic+:2019:LSSTScienceDrivers}, \gls{ARRAKIHS}\footnote{Analysis of Resolved Remnants of Accreted galaxies as a Key Instrument for Halo Surveys.} \citep{Guzman+:2022:ARRAKIHS}, and the \textsl{Nancy Grace Roman Space Telescope} \citep{Spergel+:2015:WideFieldInfrarRedSurvey} will instead provide large stream catalogs. %
    Any one projected track typically constrains its host's shape weakly, so the statistical power comes from measuring many systems. %
    We therefore need a homogeneous method that works on projected tracks, constrains each system separately, and combines these individual constraints at the population level without assuming that each halo is aligned with its galaxy's baryons. %
    Stream catalogs spanning a wide range of host masses, morphologies, and redshifts are now arriving from \gls{des}, the Stellar Streams Legacy Survey, and other facilities \citep{Miro-Carretero+:2024:ExtragalacticStellarTidal,Miro-Carretero+:2025:ExtragalacticStellarTidal,Martinez-Delgado+:2025:StellarTidalStreams, Sola+:2025}. %
    Automated pipelines can scale detection to the full survey footprint \citep{Walmsley+:2023:ZoobotAdaptableDeep,Gordon+:2024:UncoveringTidalTreasures}. %
    With track fits alone, \citet{Starkman+:2026:EuclidQ1} individually and jointly constrained host-halo shapes from 13 streams in \euclid{} Q1 imaging. %

    % Modeling trade-off
    Existing methods trade fidelity against assumptions, from full $N$-body models down to orbit fits. %
    Full $N$-body forward modeling captures tidal disruption in detail but requires expensive computation and strong priors on the progenitor and orbital history \citep{Fardal+:2013:InferringAndromedaGalaxys}; %
    particle-spray models reduce that cost at the price of more prescriptive progenitor and tidal-release assumptions \citep{Fardal+:2015:GenerationMockTidal,Pearson:2022:MappingDarkMatter}, and with GPU acceleration can search over halo parameters for external galaxies \citep{Nibauer+Pearson:2026:TestingDarkMatter,Chemaly+:2026:HierarchicalBayesianInference}. %
    Orbit-fitting approaches are cheaper still \citep{Walder+:2024:ProbingDarkMatter}, but identify the stream with an orbit, a known source of bias \citep{Sanders:2013:StreamorbitMisalignmentDangers} that angle-action modeling of the full stream distribution removes \citep{Sanders:2013:StreamorbitMisalignmentII}. %
    The curvature-based method of \citet{Nibauer+:2023}, hereafter \citetalias{Nibauer+:2023}, requires far fewer assumptions, inferring the potential's shape from projected curvature, since a stream's curvature points within \qty{90}{\degree} of the projected acceleration. %
    This method is applicable to large samples of stellar streams. %
    \citet{Wu+:2026:PotamidesMappingDark} applied it to nearby galaxies, and \citet{Starkman+:2026:EuclidQ1} to deeper \euclid{} imaging. %
    Both infer halo shapes ranging from spherical to \gls{CDM}-consistent flattening, using samples of tens of streams, as does the population-level particle-spray modeling of \citet{Chemaly+:2026:ConstraintsPopulationLevel}. %

    % Convexity as a test.
    Stream-based methods that assume a specific host potential and a prescription for the progenitor's structure and history are informative on a per-system basis, but at great risk of systematic errors. %
    Large stream catalogs make the opposite end of that trade viable, where a diagnostic assumes neither a specific potential nor a stream model and is informative only in aggregate. %
    We develop that test of galactic gravitational potentials based on the purely geometric diagnostic of the detection, or non-detection, of stellar stream segments that exhibit curvature away from the galaxy's projected \gls{CoM}. %
    Geometrically, convexity of the curvature (truly non-concavity: a straight track, or a convex one curving away from the \gls{CoM}) is the projection of the orbit's torsion, where an aspherical potential twists the intrinsic track out of the plane containing the \gls{CoM}. %
    This extends the torsion diagnostic of \citet{Bariego-Quintana:2024:TorsionStellarStreams} from 3D in the Galaxy to 2D in extragalactic systems. %
    A track concave toward the \gls{CoM} is consistent with axisymmetry about the \gls{LoS}; a convex track rules it out. %

    % The population statistic
    Because convexity provides a one-sided constraint for an individual stream, we use the diagnostic statistically: we predict the convexity rate in \gls{CDM} and \gls{SIDM} halo populations and determine how many streams are needed to distinguish between the predictions. %
    The fraction of a halo's streams with convexities decreases as the halo becomes rounder, and is zero for a spherical halo. %
    This fraction therefore traces host-halo shape, and \gls{CDM} and \gls{SIDM} predict different halo-shape distributions. %
    The convexity rate reflects both halo shape and the model-dependent orbital histories of stream progenitors \citep{Nadler+:2020:SignaturesVelocitydependentDark}. %

    % Survey readiness
    Each stream contributes a yes-or-no observation of whether it contains a convex segment. %
    Together, these observations make the catalog a survival test rather than a fit: a dark-matter model must account for the observed number of streams with convex segments. %
    The diagnostic needs only the projected track and the projected \gls{CoM}, so it suits large and heterogeneously-processed subsamples from \euclid, \gls{Rubin}, and related surveys. %

    % Organization
    We organize the paper as follows. %
    \autoref{sec:observables} motivates the diagnostic with a mock observation of GD-1 and builds the curvature-alignment statistic from the projected on-sky geometry. %
    \autoref{sec:diagnostic} derives when orbits can be convex and carries the criterion to streams. %
    \autoref{sec:triaxial_potentials} applies it to constrain individual potentials, confirming the halo triaxiality bounds numerically against particle-spray and $N$-body streams. %
    \autoref{sec:cosmology} builds population-level hypothesis tests and derives the sample sizes needed to distinguish \gls{CDM} from \gls{SIDM} at survey scale. %
    \autoref{sec:discussion} covers observational implications and \autoref{sec:conclusions} summarizes. %

% sec:intro (end)

\section{Observing Streams Around Galaxies}\label{sec:observables}

    % Roadmap
    We begin with a mock observation of GD-1 \citep{Grillmair+Dionatos:2006:DetectionColdStellar}, a stellar stream whose well-measured track we can reproject as if viewed from outside the Galaxy \citep{Malhan+:2018,Starkman+:2025}. %
    At some viewing angles the projected GD-1 stream appears convex. %
    We use this behavior to motivate the diagnostic we develop in the following sections. %

    \subsection{Mock-Observing GD-1}\label{sec:observables:gd1}

        \begin{figure}[htbp]
            \hspace{-20pt}
            \includegraphics[width=1.07\columnwidth]{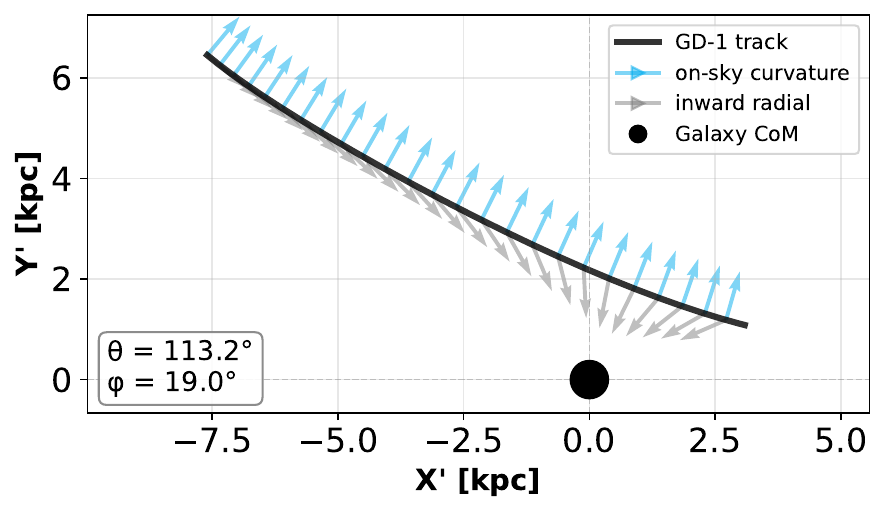}%
            \caption{%
                A mock observation of the GD-1 stream: the stream's ridge line (black) reprojected onto the sky plane as if viewed from outside the Galaxy, at the viewing angle that maximizes the convex fraction of the track. %
                Cyan arrows show the on-sky curvature direction; gray arrows show 3D curvature, which points toward the projected \gls{CoM} (black dot at the origin). %
                The curvature points away from the \gls{CoM} along the entire track; the track is everywhere convex. %
                This convexity is inconsistent with any static, centrally-concentrated potential axisymmetric about the viewing direction, including all spherical potentials. %
            }
            \label{fig:gd1_convexity}
        \end{figure}

        Before developing any formalism, we show that stream convexities exist in our own Galaxy. %
        We place an observer at infinity, take the \citet{PriceWhelan+Bonaca:2018:OffBeatenPath} track from the \package*{galstreams} package \citep{Mateu:2023:GalstreamsLibraryMilky}, and project it onto the sky plane. %

        We project GD-1 along \input{output/gd1/n_views.tex} isotropically distributed lines of sight. %
        Of these projections, \input{output/gd1/frac_any_convex.tex}\% contain a convex segment, \input{output/gd1/frac_det_ell_min.tex}\% contain one longer than \input{output/cosmology/ell_min.tex}, and \input{output/gd1/frac_fully_convex.tex}\% are convex along the full track. %
        For illustration, \autoref{fig:gd1_convexity} shows the view from Galactocentric \gls{LoS} $(\theta,\phi)=(\input{output/gd1/theta.tex},\input{output/gd1/phi.tex})$, where $\theta$ is measured from the north Galactic pole and $\phi$ from the Sun-to-center direction. %
        In this projection, GD-1 is well separated from the projected \gls{CoM}, and its curvature points away from the Galactic center along \input{output/gd1/convex_frac.tex}\% of its arc length. %
        The cyan arrows show this projected curvature, while the gray arrows show the inward radial direction. %
        The two point in opposite directions even though the track curves toward the Galactic center in 3D. %
        This convexity is plain by eye. %
        As we show below, a convex segment rules out any static, centrally concentrated potential that is axisymmetric about the \gls{LoS}, including a spherical potential. %
        This test is a projected counterpart to the torsion diagnostic of \citet{Bariego-Quintana:2024:TorsionStellarStreams}, which uses three-dimensional stream tracks to detect asphericity. %
        The result for GD-1 is unsurprising because the Milky Way is neither spherical nor static: it contains a disk and bar \citep{BlandHawthorn+Gerhard:2016:GalaxyContextStructural}, has a triaxial dark matter halo \citep{Law+Majewski:2010:SagittariusDwarfGalaxy,VeraCiro+Helmi:2013:ConstraintsShapeMilky,Han+:2022:StellarHaloGalaxy}, and moves in response to the Large Magellanic Cloud \citep{GaravitoCamargo+:2021:QuantifyingImpactLMC}. %
        The example nevertheless demonstrates that convex stream segments occur and can be identified from projected tracks. %

    % sec:observables:gd1 (end)

    \begin{figure*}[ht]
        % \centering
        \hspace{-10pt} \includegraphics[width=1.03\linewidth]{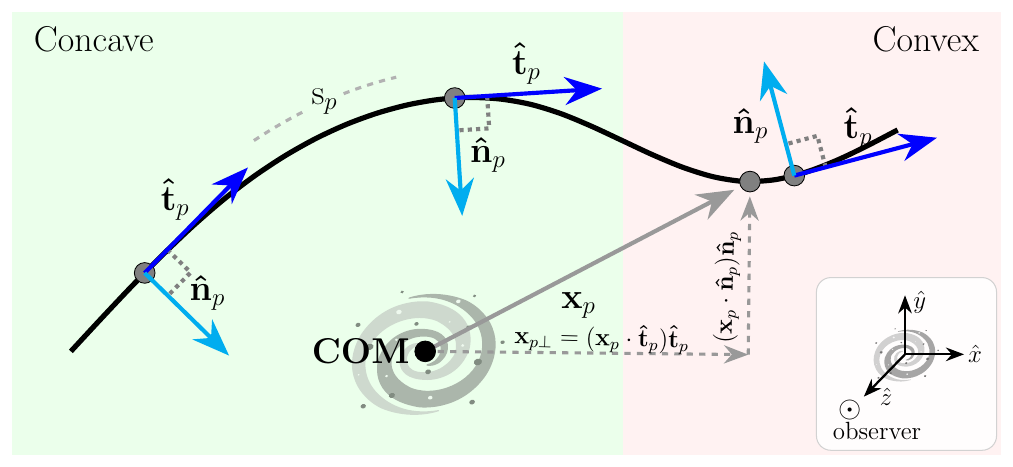}
        \caption{%
            Geometry of the on-sky curvature alignment diagnostic, shown in observer (sky-plane) coordinates. %
            The black curve is an observed projected path $\xsky(\ssky)$ of a stellar stream, parametrized by projected arc length $\ssky$ (gray dashed indicator). %
            A marker shows the location of the galaxy's projected stellar \gls{CoM}. %
            Blue and cyan arrows mark the unit tangent $\tphat$ and unit normal $\nphat$, respectively, at points along the curve. %
            The gray arrow $\xsky$ runs from the \gls{CoM} to a point on the curve and decomposes into its tangential $(\xsky\cdot\tphat)\tphat$ and normal $(\xsky\cdot\nphat)\nphat$ components (dashed). %
            The sign of $\xsky\cdot\nphat$ defines the alignment $\mathcal{A} \equiv -\rm{sign}(\nphat\cdot\xsky)$ (\cref{eq:theory:path:alignment_def}): where $\nphat$ points toward the \gls{CoM}, the curve is concave ($\mathcal{A}=+1$, green); where it points away, the curve is convex ($\mathcal{A}=-1$, red). %
            \label{fig:observables:graphic}
        }
    \end{figure*}

    \subsection{The Observer}\label{sec:observables:observer}

        % What the observer measures
        To turn an observed convexity into a constraint on the host potential, we work in terms of what an observer can measure. %
        Wide-field imaging surveys such as \euclid, \rubinlsst{}, and \gls{Roman} measure detailed on-sky positions; host redshifts are often available, but proper motions are not. %
        Stream kinematics have been measured for only a handful of systems: the M31 Giant Stream \citep{Ibata+:2004:TakingMeasureAndromeda}, the NGC~4449 stream \citep{Toloba+:2016:NewSpectroscopicTechnique}, and a disrupting dwarf in NGC~253 \citep{Toloba+:2016:TidallyDisruptingDwarf}. %
        Methods combining deep imaging with planetary-nebula kinematics may extend such measurements \citep{Valenzuela+:2026:DeepImagingMeets}, but on-sky positions remain the primary observable for large samples. %
        We therefore work in an observer's frame (\autoref{fig:observables:graphic}), measuring primarily the on-sky image. %

        % Observer vs intrinsic coordinates
        To describe the galaxy we define a coordinate system $\Xvec = (X, Y, Z)$ intrinsic to the inner galaxy, with origin at the \gls{CoM} of the stellar distribution and orientation aligned to the galaxy (e.g.\ $\Zhat$ parallel to a disk's normal). %
        The relevant components are the inner galaxy (e.g., a baryonic disk), the dark matter halo, and any massive satellite systems. %

        % Viewing coordinates
        The observer lies on the $+\zhat$ axis of a \emph{viewing} frame $(x,y,z)$ (sky plane $(x,y)$), related to the intrinsic frame by the general orthogonal transformation $ \vecC*{x,y,z} = \mbf{Q}\,\vecC*{X,Y,Z} \;$. %
        $\mbf{Q}$ decomposes as (i) a rotation mapping $\hat{\mbf{Z}}\!\to\!\zhat$, (ii) a sky-plane position-angle rotation about $\zhat$, and (iii) an optional discrete mirror $(x,y)\to(x,-y)$. %

        % Projected CoM as origin
        For any viewing-frame vector $\mbf{u}=(u_x,u_y,u_z)$ we define the sky-plane projection $\project{\mbf{u}} \equiv \vecC*{u_x,u_y}$. %
        We also use $(\cdot)_p$ for sky-plane quantities, which are not always simply projections from 3D. %
        The sky-plane \gls{CoM} is $\xcom \equiv \project{\bar{\mbf{x}}} = {M_{\mrm{tot}}}^{-1} \int_V \project{\xvec}\,\rho(\xvec,t)\,\dif<3>{\xvec}$, i.e.\ the mass-weighted mean of $\xsky=(x,y)$, marked in \autoref{fig:observables:graphic}. %

    % sec:observables:observer (end)

    \subsection{The Observed Curve}\label{sec:observables:path}

        % The stream track
        \autoref{fig:observables:graphic} shows the geometry of an observed track curve. %
        We develop it for a general curve here and apply it to an orbit in \autoref{sec:diagnostic} and to a stream ridge line in \autoref{sec:streams:match_orbit}. %
        We parametrize the path by the planar arc length $\ssky$: $\xsky(\ssky) = \vecC*{x(\ssky), y(\ssky)}$. %
        The unit tangent to the planar path is %
        \begin{equation} \label{eq:theory:path:tangent_2d}
            \tphat(\ssky) = \deriv[\ssky]{\xsky}, \qquad \|\tphat\| = 1.
        \end{equation}

        % Planar curvature and the normal
        The planar curvature $\curvaturesky(\ssky)$ measures how rapidly the planar tangent direction changes along the planar path: %
        \begin{align} \label{eq:theory:path:curvature_normal_2d}
            \deriv[\ssky]{\tphat} \;=\; \curvaturesky(\ssky)\,\nphat(\ssky),
            &\qquad
            \curvaturesky(\ssky) = \left\| \deriv[\ssky]{\tphat} \right\|,
            \\
            \|\nphat\| = 1,
            &\quad
            \tphat \cdot \nphat = 0 \, \nonumber.
        \end{align}
        where $\nphat$ is the unit normal to the planar path in the sky plane.

        % Curvature alignment
        To relate the curvature normal direction to the projected \gls{CoM}, we define a dimensionless curvature alignment diagnostic for any planar curve. %
        \begin{equation}\label{eq:theory:path:alignment_def}
            \mathcal{A} \equiv -\rm{sign}\left(\nphat\cdot \xsky\right). %
        \end{equation}
        The colored regions in \autoref{fig:observables:graphic} show this: $\mathcal{A}=+1$ means the curvature normal points ``inward'' toward the stellar \gls{CoM} $\xcom$, labeling the curve as concave (green region). %
        Conversely, $\mathcal{A}=-1$ means it points ``outward,'' labeling the curve as convex (red region). %
        The diagnostic is undefined when $\xsky=\mbf{0}$ (the galaxy's \gls{CoM}), when the curve is entirely radial, or when $\curvaturesky=0$ (an inflection point or locally straight segment). %
        In the last case we extend the classification by continuity. %
        A straight segment does not curve toward the \gls{CoM}, so we assign it to the convex class ($\mathcal{A} \leq 0$). %
        A straight segment that is itself radial falls under both cases, where the radial classification takes precedence and $\mathcal{A}$ remains undefined (\autoref{sec:diagnostic}). %

    % sec:observables:path (end)

% sec:observables (end)

\section{The Stream Convexity Diagnostic}\label{sec:diagnostic}

    In this section we develop the convexity diagnostic as a test of the potential. %
    We first determine, directly from the projected equations of motion, when a potential can host orbits with convex-in-projection segments ($\mathcal{A} \leq 0$). %
    We then specialize it to \gls{LoS}-axisymmetric centrally-concentrated potentials, and carry the orbit-level criterion onto the observable stream ridge line. % 

    % Observer-aligned cylindrical coordinates
    We work in the observer-aligned viewing frame centered on the galaxy's \gls{CoM}, illustrated in \autoref{fig:observables:graphic} and described in \autoref{sec:observables:observer}. %
    We identify this stellar \gls{CoM} with the center of the potential, the origin for the radial force $F_R = \pderiv**[R]{\Phi}$. %
    The dark matter center is expected to coincide closely with the stellar \gls{CoM}: $>$\qty{95}{\percent}  of \emph{Eagle} galaxies have offsets below \qty{700}{\parsec} \citep{Schaller+:2015:OffsetsGalaxiesTheir}, Milky-Way analogs have offsets of a few hundred parsec \citep{Kuhlen+:2013:OFFCENTERDENSITYPEAK}, and cluster brightest-galaxies have median offsets of $\sim$\qty{1}{\kilo\parsec} \citep{Roche+:2024:BrightestClusterGalaxy}. %
    Assuming the stellar \gls{CoM} is the potential center is the one assumption we make about the potential and is weaker than adopting a halo profile or a shape. %
    Also, while we defer this to a future work, jointly exploring the potential center with the data is a straightforward extension to this paper (\autoref{sec:discussion:assumptions}). %
    In this view frame, in cylindrical coordinates $(R, \theta, z)$, with $z$ along the \gls{LoS}, the projected orbit is the planar curve $(R(\theta), \theta)$. %

    % Projected equations of motion
    For an orbit in a potential $\Phi(R,\theta,z)$ the projected equations of motion are
    \begin{equation} \label{eq:theory:potential_convexities:eom}
        \ddot{R} - R\dot{\theta}^2 = -\pderiv**[R]{\Phi}, \qquad
        2\dot{R}\dot{\theta} + R\ddot{\theta} = -\frac{1}{R}\pderiv**[\theta]{\Phi},
    \end{equation}
    where we will interchangeably notate $\pderiv**[x]{}$ and $\pderiv[x]{}$.
    The projected orbit is the plane curve $R(\theta)$ traced by $(R(t), \theta(t))$. %
    \cref{eq:theory:potential_convexities:eom} is exact since the cylindrical
    decomposition separates cleanly in $z$. %

    % Signed polar curvature
    From \cref{eq:theory:path:curvature_normal_2d}, the curvature of a polar curve $R(\theta)$ is %
    \begin{equation} \label{eq:kappa_polar}
      \curvaturesky = \left| \frac{R^2 + 2R'^2 - R R''}
                       {\bigl(R^2 + R'^2\bigr)^{3/2}} \right|,
    \end{equation}
    where
    \begin{equation}
        R' \equiv \pderiv**[\theta]{R}, \; R'' \equiv \pderiv**[\theta]<2>{R}. %
    \end{equation}
    The argument of the norm is already a signed quantity. %
    Matching the sign of \cref{eq:theory:path:alignment_def}, we define
    \begin{equation}
        \curvaturesky^{\pm} \equiv \mathcal{A} \; \curvaturesky,
    \end{equation}
    where $\curvature^{\pm} > 0$ for concave segments and $\curvature^{\pm} \leq 0$ for convex ones. %
    Since \cref{eq:kappa_polar} is a geometric property of the curve $R(\theta)$, it requires no assumption about the direction of the orbit. %

    % Curvature numerator
    Differentiating $R' = \dot{R}/\dot{\theta}$ gives
    \begin{equation} \label{eq:Rpp_def}
      \dot{\theta}^3 \, R'' = \ddot{R}\,\dot{\theta} - \dot{R}\,\ddot{\theta}. %
    \end{equation}
    Substituting \cref{eq:Rpp_def} and the equations of motion (\cref{eq:theory:potential_convexities:eom}) into the numerator of \cref{eq:kappa_polar}, leaves
    \begin{equation} \label{eq:numerator}
      R^2 + 2R'^2 - RR''
        = \frac{1}{\dot{\theta}^2}
          \left(R\,F_R - R'\,F_\theta \right),
    \end{equation}
    where $F_\theta = R^{-1} \pderiv**[\theta]{\Phi}$. %
    Since $\dot{\theta}^2 > 0$, the projected orbit is convex if and only if locally %
    \begin{equation} \label{eq:concavity}
        \boxed{
            R\;F_R \; \leq \; \pderiv[\theta]{R} \;F_\theta,
        }
    \end{equation}
    with strict inequality giving convexity and equality giving a straight segment. %
    The left side of \cref{eq:concavity}, $R\, F_R$, is the radially-weighted radial force. %
    For mass profiles including a Keplerian potential, it pulls the track concavely toward the \gls{CoM}. %
    The right side is the azimuthal force, weighted by the local projected slope $R'$, partly set by how plunging the orbit is. %
    Where the projected track is nearly radial, e.g. for an eccentric, radially plunging orbit, the azimuthal term is amplified, so even a weak azimuthal perturbation can produce a convexity. %

    In 3D, projected convexity can come from torsion. %
    In an aspherical potential orbits experience nonzero torsion, which twists their osculating planes out of the plane containing the \gls{CoM} \citep{Bariego-Quintana:2024:TorsionStellarStreams}. %
    Projected onto the sky, this torsion can reverse the sign of the apparent curvature. %
    Torsion gives the 3D picture, but we work from the projected equations of motion to derive the convexity threshold explicitly. %

    % Curvature sign from the projected acceleration
    Equivalently, \cref{eq:concavity} can be interpreted as a condition on the projected acceleration along the path. %
    That acceleration is the gravitational field $\asky = -F_R\,\hat{R} - F_\theta\,\hat{\theta}$. %
    Resolving $\asky$ relative to the projected velocity $\vsky$ (speed $\speedsky \equiv \|\vsky\|$), only its perpendicular part ${\asky}_{\perp} \equiv \asky - (\asky\cdot\tphat)\tphat$ bends the track. %
    This perpendicular acceleration fixes the curvature normal through $\curvaturesky\,\nphat = {\asky}_{\perp}/\speedsky^2$. %
    Substituting into \cref{eq:concavity}, the kinematic prefactors cancel and the convexity condition reduces to the sign of the projected acceleration toward the \gls{CoM}, %
    \begin{equation}\label{eq:theory:path:orbit:alignment_from_a}
        \mathcal{A}
        = -\mathrm{sign}\left({\asky}_{\perp}\cdot\xsky\right). %
    \end{equation}
    The criterion depends on $\asky$ at that location, not on the speed $\speedsky$. %
    This speed-independence underlies the curvature method of \citetalias{Nibauer+:2023}: the shape of the projected curve alone constrains the projected acceleration field. %
    With the direction-independence of \cref{eq:kappa_polar}, we need only on-sky positions. %

    % LoS axisymmetry
    For a potential axisymmetric about the \gls{LoS}, $\pderiv**[\theta]{\Phi} = 0$ in observer-aligned frames, so the azimuthal-force term on the right-hand side of \cref{eq:concavity} vanishes. %
    The convexity condition then reduces to $F_R \leq 0$, a net outward radial force. %
    Centrally-concentrated potentials, such as the \gls{NFW} profile \citep{Navarro+:1996:StructureColdDark,Navarro+:1997:UniversalDensityProfile}, have $\pderiv**[R]{\Phi} > 0$ everywhere except at the \gls{CoM}, so nothing satisfies this condition. %
    Therefore, in any \gls{LoS}-axisymmetric centrally-concentrated potential, the projected orbit is always concave toward the \gls{CoM}, independent of inclination and eccentricity. %
    The exceptions are when $R=0$ or $\dot\theta=0$, where $\curvature$ (\cref{eq:theory:path:alignment_def}) is undefined. %
    A convex segment therefore requires an azimuthal force in the observer's frame. %
    This force can be intrinsic (e.g.,\ a triaxial halo or a bar) or induced by off-axis viewing an axisymmetric potential. %

    % The edge-on degeneracy
    The diagnostic fails in only one case, a {tangent-passage}, where the projected tangent points straight at the projected \gls{CoM} ($\xskyperp = \mbf{0}$) and $\mathcal{A}$ vanishes. %
    The angular momentum about the \gls{LoS}, $L_{\mrm{los}} = R^2\dot{\theta}$, sets how close the tangent line comes to the \gls{CoM}: the perpendicular distance is $\|\xskyperp\| = R^2|\dot{\theta}|/\speedsky = |L_{\mrm{los}}|/\speedsky$. %
    A tangent-passage therefore means $L_{\mrm{los}} = 0$. %
    In a \gls{LoS}-axisymmetric potential $L_{\mrm{los}}$ is conserved, so this is all-or-nothing: either the orbit projects to a radial line through the \gls{CoM} and $\mathcal{A}$ is undefined along its whole length, or the tangent line never reaches the \gls{CoM} and $\mathcal{A}$ is defined everywhere. %
    The first case looks like a straight track pointing at the \gls{CoM}, and we exclude such tracks from the diagnostic. %
    It is the only straight projected track allowed by \gls{LoS}-axisymmetry, and so the only exception to the straight-segment convention of \autoref{sec:observables:path}. %

    \subsection{Applying the Diagnostic to Streams}\label{sec:streams:match_orbit}

        % From orbit to track
        The criterion so far applies to an orbit, but a stream is not an orbit. %
        It is a collection of stars stripped at different times onto nearby orbits, and we observe the ridge line they trace (\autoref{fig:observables:graphic}). %
        We evaluate the diagnostic on that stream ridge line, so we need to know when the ridge line and the orbits composing it have the same concavity. %

        % The mismatch angle
        Let $\tphat^{\mrm{str}}$ be the stream (ridge-line) tangent (\cref{eq:theory:path:tangent_2d}) and $\tphat^{\mrm{orb}}$ the tangent of the orbit of the star at that point. %
        The angle between them is %
        \begin{equation}\label{eq:streams:match_orbit:tangent_mismatch_angle}
            \delta\theta(\ssky)
            \equiv
            \arccos\!\left(\bigl|\tphat^{\mrm{str}}(\ssky)\cdot\tphat^{\mrm{orb}}(\ssky)\bigr|\right). %
        \end{equation}
        \Cref{eq:theory:path:curvature_normal_2d} fixes the normal to the tangent, so the same angle separates the two curvature normals. %
        Where $\delta\theta$ is small, the ridge line and the orbit have the same $\mathcal{A}$. %

        % Tangent spread across the width
        How fast the stream widens sets that angle $\delta\theta$. %
        Let $W(\ssky)$ be the root-mean-square (RMS) transverse spread of stars about the ridge line, set by the release-velocity dispersion smearing the debris into a band \citep{Kupper+:2010:TidalTailsStar,Kupper+:2012:MoreStructureTidal,Bovy:2014:DynamicalModelingTidal}. %
        Over a short arc length $\delta\ssky$ the ridge line is locally straight, but its edge moves by $\delta W$, so the edge crosses the ridge direction at an angle $\tan\delta\theta = \delta W/\delta\ssky$. %
        Where this angle is large the orbits point, and so curve, away from the ridge line. %
        Therefore, even for relatively wide streams, it is where the width changes rapidly that the ridge line and orbits can have different concavities; elsewhere they agree. %

        The stream is within the transverse offsets $\pm W(\ssky)$ from the ridge line, and has slopes relative to the ridge line bounded by $|\pderiv**[\ssky]{W}|$. %
        The tangent mismatch therefore satisfies %
        \begin{equation}\label{eq:streams:width_bound}
            \delta\theta(\ssky)
            \;\lesssim\;
            \bigl|\pderiv**[\ssky]{W(\ssky)}\bigr| . %
        \end{equation}
        \autoref{sec:ridgeline_robustness:width_bound} derives similar bounds for every track within the stream, not only for its edges. %
        The tangent mismatch therefore depends on the width gradient more than the width itself: even a wide stream has well-aligned tangents where its width is constant. %

        % Tangent passages
        Whether $\delta\theta$ flips the sign of $\mathcal{A}$ also depends on $\|\xskyperp\|$, the perpendicular distance from the projected \gls{CoM} to the tangent line. %
        \Cref{eq:theory:path:alignment_def} defines the alignment; rotating the normal by $\delta\theta$ changes that product by $\|\xsky\|\sin\delta\theta$, so the sign holds only where %
        \begin{equation}\label{eq:streams:match_orbit:sign_robust}
            \|\xskyperp\|
            \;\gg\;
            \|\xsky\|\,\sin\delta\theta . %
        \end{equation}
        The left-hand side is small where the track points near the \gls{CoM}. %
        This is the tangent-passage of \autoref{sec:diagnostic}, where even small $\delta\theta$ can reverse $\mathcal{A}$. %
        We call the band around it in which \cref{eq:streams:match_orbit:sign_robust} fails the \emph{near-tangent strip}, and exclude it from the diagnostic. %
        With \cref{eq:streams:width_bound}, the ridge-line alignment matches that of its constituent orbits where %
        \begin{equation}\label{eq:streams:match_orbit:transfer}
            \|\xskyperp\| \gg \|\xsky\|\,\bigl|\pderiv**[\ssky]W(\ssky)\bigr|.
        \end{equation}
        Every quantity is measurable from imaging and does not require assumptions relating the width to the orbital scale. %
        Where the width varies by $|\pderiv**[\ssky]{W}| \sim W/\|\xsky\|$, the condition becomes $\|\xskyperp\| \gg W$, i.e.\ the projected \gls{CoM} must lie outside a strip of width comparable to the stream width from the local tangent line. %
        These derivations are geometric and independent of how the debris was released, relying instead on continuity and coherence arguments intrinsic to non-disrupting streams. %

        % Numbers and where it holds
        Empirically, GD-1 and Pal~5 have RMS widths $W\sim \qtyrange{50}{150}{\parsec}$ over tens of \kpc\ \citep{Koposov+:2010:ConstrainingMilkyWay, Erkal+:2017:SharperViewPal5, Bonaca+:2020:VariationsWidthDensity}, and dwarf-galaxy streams reach $W\sim \qtyrange{2}{3}{\kilo\parsec}$ (Sagittarius, \citealt{Correnti+:2010:NorthernWrapsSagittarius}; Centaurus~A, \citealt{Pearson:2022:MappingDarkMatter}) over comparable or greater lengths, giving $|\pderiv**[\ssky]{W}|\sim W/L\lesssim 10^{-2}$. %
        The excluded near-tangent strip, $|\xskyperp| \lesssim W$, is therefore \numrange{10}{100} times wider for dwarf-galaxy streams than for globular-cluster streams, so for the latter the diagnostic applies closer to edge-on viewing angles and to more radial projected tracks. %
        Globular-cluster streams are the preferable system, but very few are known beyond the Local Group \citep{KielHolm+:2026:EvidenceForFirstGCStellarStream} and the broader dwarf-galaxy streams dominate extragalactic catalogs. %
        The curvature criterion of \citetalias{Nibauer+:2023} requires a similar agreement between tracks and orbit, there emphasizing coherence as a survival condition of the stream to bound the track-orbit angles. %
        Here we further require the angular offsets not to change $\mathcal{A}$ between track and orbits. %
        Differences of order the stream width therefore do not change the sign of $\Astr$ (\autoref{sec:ridgeline_robustness:smoothing}). %
        Where $|\pderiv**[\ssky]{W}|$ is large $\Astr$ can become unreliable: in fanning regions near apocenter or after a strong tidal interaction, and in the shells and caustics formed by low-angular-momentum debris \citep{Hendel:2015:TidalDebrisMorphology}. %
        For an example in the Milky Way, parts of Pal~5's thin track lie in a fanning region \citep{Starkman+:2020:ExtendedPal5Stream, Pearson+:2015:TidalStreamMorphology} and would be unusable. %
        Also, within a few tidal radii of an intact progenitor $\Phi_{\mrm{prog}}$ will shift the curvature normal. %
        All three are identifiable in imaging, so we should evaluate $\Astr$ only on segments well separated from the progenitor where $W(\ssky)$ varies slowly. %
        For the rest of this work we therefore apply the orbit criterion of \cref{eq:theory:path:orbit:alignment_from_a} directly to observed stream tracks, with $|\pderiv**[\ssky]{W}|$ as the check on where it applies. %

    % sec:streams:match_orbit (end)

% sec:diagnostic (end)

\section{Convexities in Triaxial Potentials}\label{sec:triaxial_potentials}

    % Roadmap
    Convex segments cannot occur in a \gls{LoS}-axisymmetric centrally-concentrated potential, which we prove analytically in \autoref{sec:triaxial_potentials:axisymmetric}. %
    Convex segments can occur in a triaxial potential from certain viewing directions, which we establish using numerical orbits and the general form of $\mathcal{A}$ in \autoref{sec:triaxial_potentials:orbits}. %
    Last, in \autoref{sec:triaxial_potentials:simulations} we confirm the results of \autoref{sec:streams:match_orbit} and show that convexities occur in simulated streams. %

    \subsection[Ruling out LoS-Axisymmetric Centrally-Concentrated Mass Profiles]{Ruling out \gls{LoS}-Axisymmetric Centrally-Concentrated Mass Profiles}\label{sec:triaxial_potentials:axisymmetric}

        % Transfer to the observed track
        In a \gls{LoS}-axisymmetric potential the field has no azimuthal component, so the projected acceleration is antiparallel to $\xsky$ and ${\asky}_{\perp} = -(\pderiv[R]{\Phi}/R)\,\xskyperp$ with $\pderiv**[R]{\Phi} > 0$ for a centrally-concentrated profile (\autoref{sec:diagnostic}). %
        \Cref{eq:theory:path:orbit:alignment_from_a} then gives
        \begin{equation}\label{eq:theory:orbit:axisymmetric:alignment}
            \mathcal{A}
            = -\rm{sign}\!\left({\asky}_{\perp}\cdot\xsky\right)
            = +1, %
        \end{equation}
        independent of inclination and eccentricity. %
        The ridge line inherits the orbit's curvature sign wherever the width bound of \autoref{sec:streams:match_orbit} holds, so this applies to the observed stream track outside the near-tangent strip. %
        Spherical potentials $\Phi(r)$ are the trivial case: every axis through the \gls{CoM} is a symmetry axis, so the inward-normal condition holds for any viewing geometry. %

        % Ruling out LoS axisymmetry
        Away from the edge-on tangent-passage geometry, which is identifiable from its straight-line morphology (\autoref{sec:diagnostic}), observation of a single convex arc ($\mathcal{A} \leq 0$) anywhere along an observed track rules out the whole class: any spherical potential, and more generally any \gls{LoS}-axisymmetric centrally-concentrated one. %

    % sec:triaxial_potentials:axisymmetric (end)

    \subsection{Orbits in Triaxial Potentials}\label{sec:triaxial_potentials:orbits}
        \begin{figure*}[ht]
            \centering
            \includegraphics[width=1\textwidth]{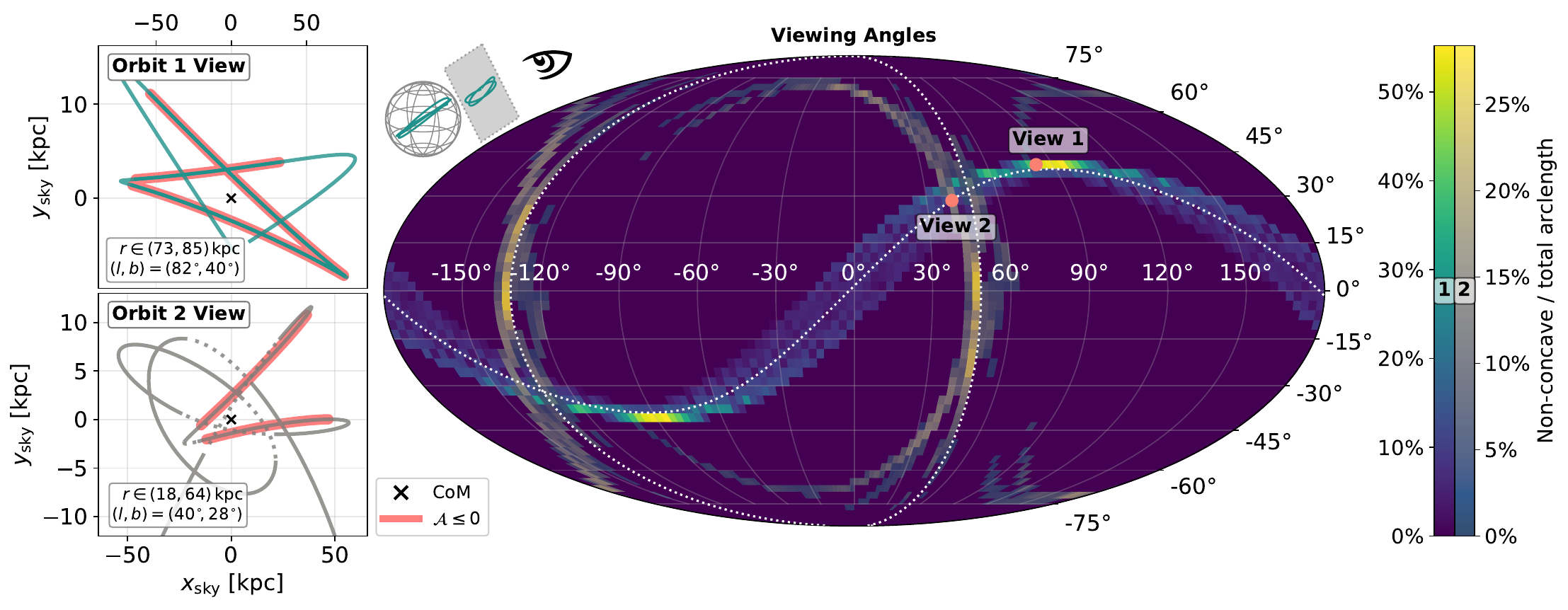}\\[1ex]
            \caption{%
                \emph{Fraction of a projected orbit that appears convex, across viewing angles, for two orbits in a triaxial \gls*{NFW} potential}. %
                \emph{Left}: sky-plane projections of Orbit~1 (top, $e\in\input{output/triaxial/orbit1_ecc.tex}$) and Orbit~2 (bottom, $e\in\input{output/triaxial/orbit2_ecc.tex}$), each shown at the viewing angle specified with a red dot in the right plot. %
                Red segments mark the convex portions; the cross denotes the \gls{CoM}. %
                Dotted segments show where Orbit~2 (never Orbit~1) passes within \input{output/triaxial/r_near.tex} of the \gls{CoM}, where baryonic contributions are more significant. %
                \emph{Right}: Mollweide projections over all viewing directions for Orbit~1 (viridis, labeled~1) and Orbit~2 (cividis, labeled~2). %
                At each pixel, the color encodes the fraction of the projected orbit that appears convex when viewed from that direction. %
                White dotted great circles mark the mean osculating plane of each orbit; viewing directions lying on these circles see parts of the orbit edge-on. %
                The brightest regions are the viewing angles at which the largest convex fraction is visible. %
                \label{fig:orbit_triaxial}
            }
        \end{figure*}

        % Intro
        No halo is perfectly axisymmetric, so ruling out that class (\autoref{sec:triaxial_potentials:axisymmetric}) is only a starting point. %
        Here, we investigate how a triaxial potential produces convexities, deriving the alignment $\mathcal{A}$ for a general triaxial potential and showing convexities in numerical orbit solutions. %

        % Example orbits
        \autoref{fig:orbit_triaxial} shows two test-particle orbits integrated for \input{output/triaxial/t_integrate.tex} in a pure triaxial density \gls{NFW} potential ($M = \input{output/triaxial/m_tot.tex}$, $r_s = \input{output/triaxial/r_s.tex}$, $b/a = \input{output/triaxial/q1.tex}$, $c/a = \input{output/triaxial/q2.tex}$; no disk component), at low and moderate eccentricity (Orbit~1, $e \in \input{output/triaxial/orbit1_ecc.tex}$; Orbit~2, $e \in \input{output/triaxial/orbit2_ecc.tex}$). %
        The triaxial \gls{NFW} density model \citep{Navarro+:1996:StructureColdDark, Navarro+:1997:UniversalDensityProfile} is the most relevant triaxial density profile for \gls{CDM} dark matter halos. %
        We hold each orbit fixed while the viewing direction varies: for each direction we project the orbit onto the observer's sky and measure the projected arc-length fraction that is convex. %
        The left panels show each orbit as it appears from the viewing direction marked in the right panel, with the convex arcs in red. %
        The right panel maps this convex fraction over viewing direction in Mollweide projection, darkest where no segment is convex. %
        Both orbits lie predominantly beyond \input{output/triaxial/r_near.tex}, well into the dark-matter-dominated regime for galaxies. %
        The rare interior segments (dotted) do not overlap the convex arcs --- the convexities are not caused by an inner galaxy component like a disk, which this model does not include. %

        % Viewing-angle maps
        From most viewing directions the two orbits are entirely concave; at some viewing angles segments become convex. %
        The fraction of each orbit that is convex at a given viewing angle reaches $\sim \input{output/triaxial/orbit1_peak_convex.tex}\%$ for Orbit~1 and $\sim \input{output/triaxial/orbit2_peak_convex.tex}\%$ for Orbit~2. %
        Most orbit segments that become convex do so when viewed near edge-on to the orbit's osculating plane (dotted white lines). %
        Face-on views, perpendicular to the orbital plane, show almost no orbit convexities. %
        Geometrically, this is expected. %
        Torsion from the triaxial potential pulls orbits from the spherical osculating plane, and so convex deviations from great circles are largest here, where the torsion signals are most evident \citep{Bariego-Quintana:2024:TorsionStellarStreams}. %
        Within each orbit, the intersections of its great-circle bands mark viewing directions where multiple convex segments of that same orbit are simultaneously visible. %

        % Derivation
        We now build some analytic understanding of orbit convexities in triaxial potentials. %
        \autoref{fig:orbit_triaxial} showed convexities in triaxial \gls{NFW} density profiles; however from \cref{eq:concavity} we understand that convexities arise more straightforwardly from the potential. %
        As further explained in \autoref{sec:triaxiality_density_vs_potential}, triaxial densities do not generically have simple analytic parametrizations in the potential, but are nearly the same as a triaxial potential up to perturbative terms for realistic halos. %
        We therefore consider triaxial potentials. %

        With the convexity diagnostic we can build analytic understanding and constraints on triaxial potentials without ever assuming a specific potential parametrization like an \gls{NFW} model. We only require the generic form $\Phi(\xvec) = \Phi(m)$ with %
        \begin{equation}\label{eq:theory:orbit:triaxial:m_def}
            m^2 \equiv \xvec^{\rm T}\mbf{M}\xvec,
        \end{equation}
        where the symmetric shape tensor has $\mbf{M} = \mrm{diag}(a^{-2},b^{-2},c^{-2})$ in its principal-axis frame with $a \ge b \ge c$. %
        In observer-aligned coordinates $\mbf{M}$ is generally not diagonal, and
        \begin{equation}\label{eq:theory:orbit:triaxial:g}
            \mbf{g} = -\frac{\Phi'(m)}{m}\,\mbf{M}\xvec,
            \qquad \Phi'(m) \equiv \frac{\dif\Phi}{\dif m} > 0.
        \end{equation}
        With $\mbf{M}$ positive definite, $\Phi'(m)>0$ ensures $\mbf{g}$ points inward. %
        Decomposing $\xvec = \xsky + z\zhat$ and projecting, %
        \begin{equation}\label{eq:theory:orbit:triaxial:asky}
            \asky = -\frac{\Phi'(m)}{m}\bigl(\proj{\mbf{M}}\,\xsky + z\,\mbf{m}_{\zhat}\bigr), %
        \end{equation}
        where $\proj{\mbf{M}}$ is the $2\times 2$ upper-left submatrix of $\mbf{M}$ in observer-aligned coordinates, and $\mbf{m}_{\zhat} \equiv \project{\mbf{M}\zhat}$ is the sky-plane projection of the $\zhat$ column of $\mbf{M}$. %
        Together these set the decomposition $\project{\mbf{M}\xvec} = \proj{\mbf{M}}\,\xsky + z\,\mbf{m}_{\zhat}$ for any $\xvec = \xsky + z\zhat$. %
        Using the identity ${\asky}_{\perp}\cdot\xsky = \asky\cdot\xskyperp$ and substituting into \cref{eq:theory:path:orbit:alignment_from_a}, the factor $-\Phi'(m)/m < 0$ flips the overall sign, giving
        \begin{equation}\label{eq:theory:orbit:triaxial:alignment}
            \mathcal{A}
            = \rm{sign}\!\bigl[(\proj{\mbf{M}} \xsky) \cdot \xskyperp + z\,\mbf{m}_{\zhat} \cdot \xskyperp \bigr]. %
        \end{equation}
        The intrinsic term is non-zero when $\proj{\mbf{M}}\xsky$ is not parallel to $\xsky$, i.e.\ when the projected isopotentials are non-circular ($\proj{\mbf{M}} \not\propto \mbf{I}$). %
        The projective term $\mbf{m}_{\zhat}$ vanishes only when $\zhat$ is a principal axis of $\mbf{M}$. %
        Otherwise it couples the unobserved \gls{LoS} distance $z$ into the sky-plane dynamics. %
        Both terms of \cref{eq:theory:orbit:triaxial:alignment} are at work in the orbits of \autoref{fig:orbit_triaxial}: the isopotentials there are non-circular and almost every viewing direction is off a principal axis. %
        Their relative size cannot be easily inferred from the figure, since the viewing directions with convexities align more with the orbits than with the potential. %

        % From 1 halo to a population
        For a general viewing direction, the unknown \gls{LoS} distance $z$ stops us from inferring the shape of an individual halo from the projected track. %
        Generative stream models can supply a distance track, but conditional on a specific model for the potential \citep{Fardal+:2015:GenerationMockTidal,Pearson:2022:MappingDarkMatter,Nibauer+Pearson:2026:TestingDarkMatter}. %
        We instead record only whether the projected track contains a convex segment. %
        Across a population of streams, the frequency of such segments depends on the distribution of halo shapes predicted by the dark matter model. %
        The convexity rate therefore tests a cosmological model without fitting the potential of each host, as we show in \autoref{sec:cosmology}. %

    % sec:triaxial_potentials:orbits (end)

    \subsection{Convexities in Simulated Streams}\label{sec:triaxial_potentials:simulations}

        \begin{figure*}[t]
            \centering
            \includegraphics[width=\textwidth]{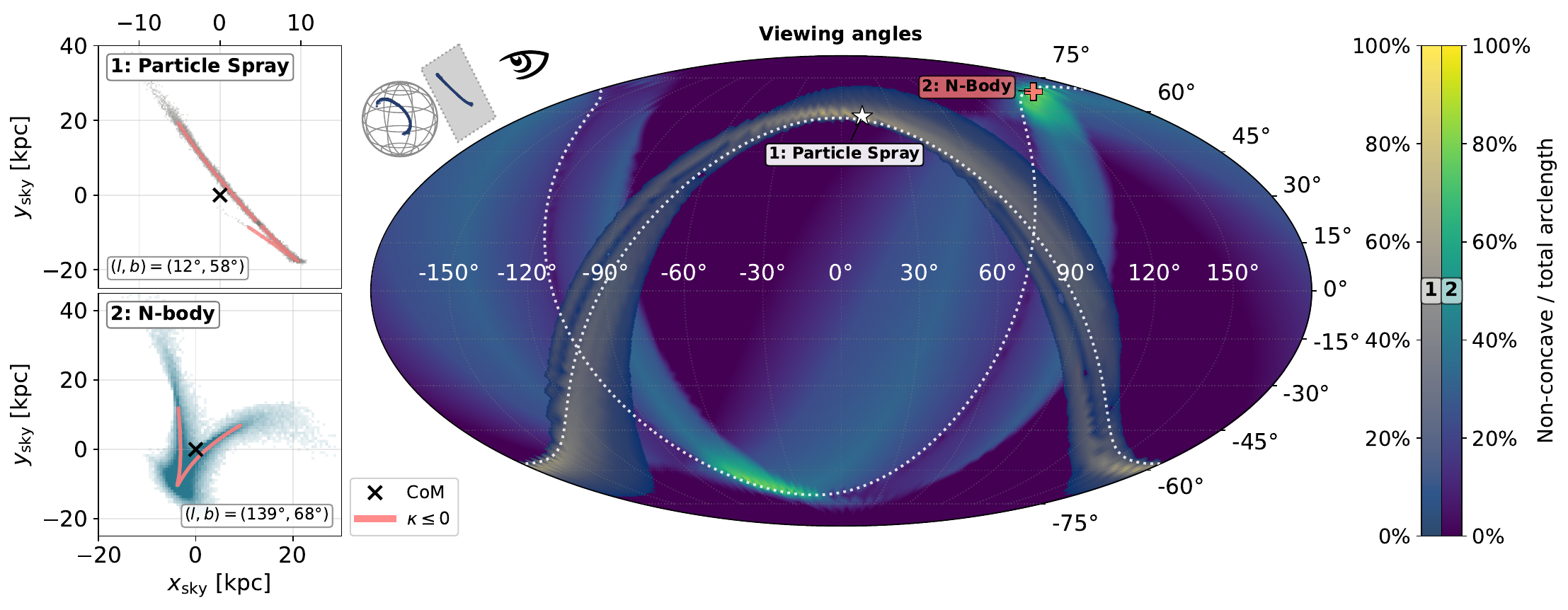}%
            \caption{
                    \emph{Convex arc-length fraction of two different kinds of simulated streams, across viewing angles}. %
                    The particle spray simulation uses the same initial condition and triaxial \gls{NFW} potential as \autoref{fig:orbit_triaxial}, while the $N$-body comes from \citetalias{Nibauer+:2023}. %
                    \emph{Left}: sky-plane projections of particle spray (top) and $N$-body (bottom), each shown at the viewing angle specified with a white star and an orange cross in the right plot. %
                    Red segments mark the convex portions; the black cross denotes the \gls{CoM}. %
                    \emph{Right}: Mollweide projections over all viewing directions for Particle spray Simulation~1 (viridis, labeled~1) and $N$-body simulation~2 (cividis, labeled~2). %
                    Panels are otherwise as in \autoref{fig:orbit_triaxial}, which shows convexities in orbits. %
                    This figure extends that result to streams, finding visible convexities in particle-spray and $N$-body streams.%
                    }
            \label{fig:simulations:spray_and_nbody}
        \end{figure*}

        % Convexities in Simulations
        We test the diagnostic on two simulated streams: one particle-spray stream and one $N$-body stream, each from a different dwarf galaxy progenitor. %
        Both confirm the analytic prediction that large-scale, $>10 kpc$ convexities occur and can be detected in the projected stream tracks. %
        \autoref{fig:simulations:spray_and_nbody} presents these streams, with the top left panel showing the particle-spray stream and the bottom left panel showing the $N$-body stream. %
        The convex portions are drawn in red, relative to the projected \gls{CoM} at the cross. %
        We simulate the particle-spray stream in the triaxial \gls{NFW} potential of \autoref{fig:orbit_triaxial}, with the same initial conditions. %
        To better match the length of a dwarf galaxy stream, we use a \qty{6.5}{\giga\year} integration time with progenitor mass $M_{\rm prog} = \qty{e6}{\Msun}$. %
        The $N$-body simulation is drawn from and described in \citetalias{Nibauer+:2023}, and differs from the particle-spray example in both its potential --- here a $q=1.5$ flattened logarithmic halo --- and its initial conditions. %
        The integration time for the $N$-body simulation is \qty{10.375}{\giga\year}. %

        % Across viewing Angles
        For each simulated stream we build a stream track and analyze that track for convexities across viewing directions. %
        To build the stream track we mix methods: doing initial ordering with \package*{potamides} \citep{Wu+:2026:PotamidesMappingDark}, then fitting and outlier removing using \package*{phasecurvefit} \citep{Starkman+2026:phasecurvefit}, and finally fitting a full 3D ridge line, \citep{Nibauer+:2022,Starkman+:2023:FastTrackStellarStreamPaths}. %
        From this 3D track we project it onto each viewing direction, and measure the convex arc-length fraction in each 2D projection. %
        The Mollweide plot of \autoref{fig:simulations:spray_and_nbody} shows the convex fraction across viewing directions. %
        The highest convex fractions concentrate near the osculating plane of the stream, consistent with the analytic prediction that convexity is most visible when the stream is viewed near edge-on in its orbital plane (\autoref{sec:triaxial_potentials:orbits}). %
        Both simulations confirm the predictions of \autoref{sec:diagnostic} and \autoref{sec:triaxial_potentials:orbits} that the convex arc-length fraction is non-zero in non-spherical potentials and goes to zero in the spherical limit. %

    % sec:triaxial_potentials:simulations (end)

% sec:triaxial_potentials (end)

\section{Results: Building a Cosmology Test}\label{sec:cosmology}

    % Intro
    A single convexity primarily tells us that one halo is not \gls{LoS}-axisymmetric. %
    Across a galaxy population, the triaxiality of the halos determines the fraction of convexity-containing streams. %
    Since halo triaxiality is a prediction of the dark matter model, so too is the convex stream fraction. %
    In this section we turn the convexity diagnostic into a test of dark matter model predictions. %
    
    % Shape -> DM Model
    \gls{CDM} predicts triaxial halos while self-interactions drive \gls{SIDM} halos toward sphericity (\autoref{sec:intro}). %
    Outside the inner halo the two can therefore have nearly identical radial density profiles but different three-dimensional shapes, even at Milky-Way masses \citep{Bautista+:2025:JeansModelShapes,Adhikari+:2025:AstrophysicalTestsDark}. %
    \gls{SIDM} shapes do gradually approach \gls{CDM} at large enough radii, as the effects of self-interactions weaken, leaving a radial window whose extent depends on the \gls{SIDM} cross-section (\autoref{sec:cosmology:microphysics}). %
    The useful radial window is outside the baryon-dominated inner halo but inside the radius beyond which \gls{SIDM} and \gls{CDM} shapes converge. %
    Within this window, the models can have similar radial density profiles but different three-dimensional shapes, and convexity measures that shape difference. %

    % Why outer streams
    Inside the halo scale radius $R_s$ --- for \gls{NFW} halos this is the radius where the logarithmic slope of the density profile is $-2$, $R_s\approx\qtyrange{15}{20}{\kilo\parsec}$ for galaxies --- baryons contribute substantially to the potential and thus also to any measured non-sphericity. %
    Several baryonic scale radii out, the bulge and disk are sub-dominant, and the halo shape reflects the cosmological triaxiality of \gls{CDM} or \gls{SIDM} \citep{Vera-Ciro+:2011:ShapeDarkMatter,Baptista+:2022:OrientationsDMHalos}. %
    The dwarf galaxy streams that \euclid, \gls{Rubin}, and \gls{Roman} will identify typically lie many scale radii away, far enough that imaging shows the separation. %
    Survey selection therefore puts these streams in the dark-matter-dominated regime. %

    % The population measurement
    After we filter out the near-tangent strip where the diagnostic is unreliable (\autoref{sec:streams:match_orbit}), each stream is just a yes-or-no measurement: does it show a convex segment? %
    In \autoref{sec:triaxial_potentials} we demonstrate that the answer is yes for a small fraction of viewing directions, in potentials motivated by typical \gls{CDM} galaxy-mass hosts \citep{Allgood+:2006:ShapeDarkMatter,Vera-Ciro+:2011:ShapeDarkMatter,Giocoli+:2026:AIDATNGProject3D}. %
    The halo-shape distribution determines the rate. %
    Therefore, for a \gls{SIDM} host of the same mass and radial profile the answer is yes less often than for \gls{CDM}. %
    Larger self-interaction cross-sections $\sigma/m$ round the halo out to larger radii, lowering the \gls{SIDM} convexity rate and widening the difference between the models (\autoref{sec:cosmology:microphysics}; \citealt{Brinckmann+:2018:StructureAssemblyHistory,Fischer+:2024:CosmologicalIdealizedSimulations}). %
    Across a catalog of $N$ streams, the number $K$ of streams with a convexity gives an observed rate $\lamObs \equiv K/N$, which the two dark matter models predict differently. %

    % Roadmap.
    \autoref{sec:cosmology:rates} motivates the two rates from simplified simulations, establishing their order of magnitude and sensitivity to halo shape. %
    We then use that catalog three ways, each through a different statistic: %
    testing whether it excludes a model outright (\autoref{sec:cosmology:excluded}, a binomial tail probability), %
    comparing the two models (\autoref{sec:cosmology:favored}, a likelihood ratio), %
    and forecasting how many streams a survey needs to reach a given confidence (\autoref{sec:cosmology:sample_size}, a power calculation). %

    \subsection{The Stream Convexity Rate}\label{sec:cosmology:rates}

        \begin{figure*}[t]
            \centering
            \includegraphics[width=\linewidth]{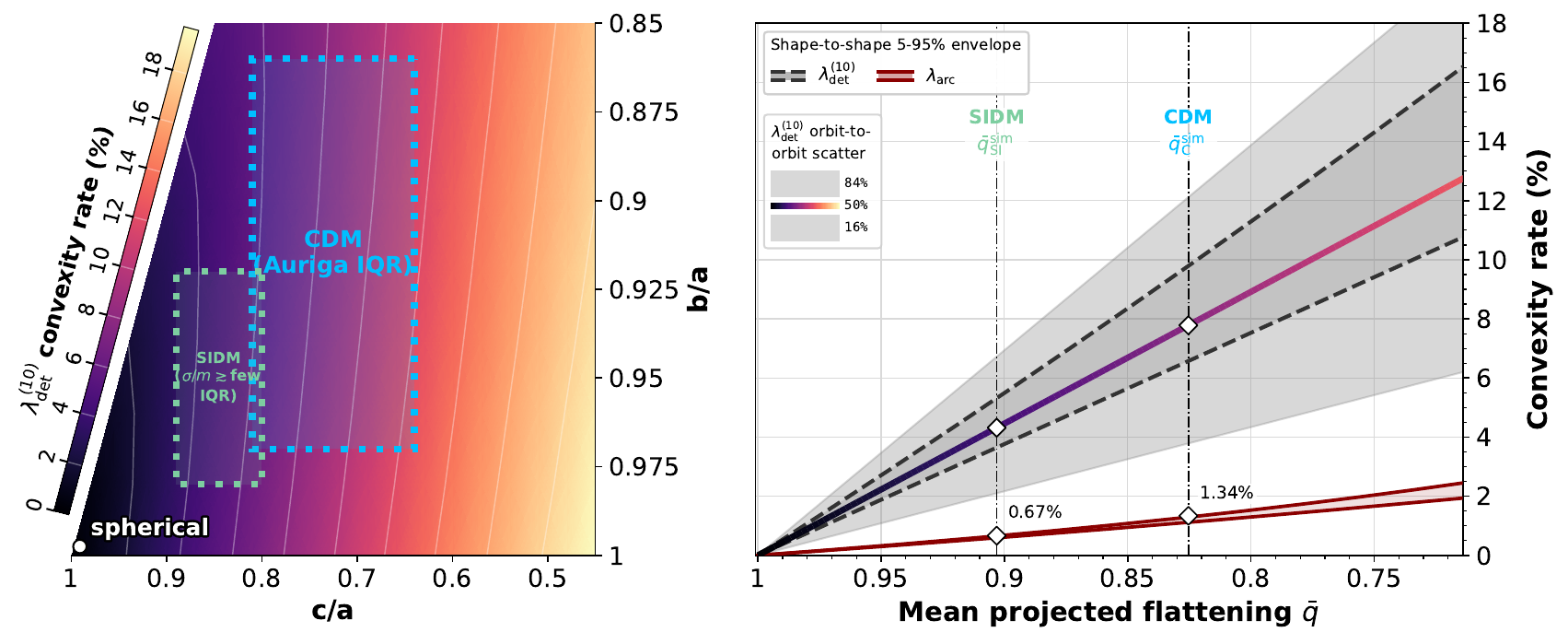}
            \caption{%
                Per-stream convexity rates in a grid of triaxial \gls{NFW} halo potentials. %
                \textit{Left:} %
                colors show $\lamdadet{10}$, the fraction of streams with a convex segment longer than $\input{output/cosmology/ell_min.tex}$. %
                White contours are lines of constant $\lamdadet{10}$ (colorbar). %
                The blank wedge has $b/a<c/a$, excluded by definition. %
                The blue dotted box is the joint interquartile region of the Auriga hydrodynamic \gls{CDM} halos in $(b/a,\,c/a)$. %
                The green dotted box shows an extrapolated with-baryons \gls{SIDM} counterpart, obtained by rounding those halo shapes for $\sigma/m\gtrsim\text{a few}\ \unit{\square\centi\meter\per\gram}$. % 
                \textit{Right:} %
                the same potentials are plotted against $\bar q$, their mean projected minor-to-major density-axis ratio. %
                The colored curve and gray band show the median and orbit-to-orbit spread of $\lamdadet{10}$; black dashed curves show how the rate varies among intrinsic halo shapes with the same $\bar q$, e.g. the difference between more oblate and more prolate halos. %
                Solid red curves and their shaded envelope show how $\lambda_{\rm arc}$ varies among intrinsic halo shapes with the same $\bar q$. %
                $\lambda_{\rm arc}$ is the fraction of stream length that is convex. %
                $\lambda_{\rm arc}$ requires convexity over much of a stream, whereas $\lamdadet{10}$ requires one resolved segment. %
                A survey's rate should lie between them. %
                Black dot-dashed lines, labeled \gls{CDM} in blue and \gls{SIDM} in green, mark the mean $\bar q$ of the two populations. %
                At each line, the upper white diamond gives the mean $\lamdadet{10}$ and the lower white diamond gives the mean $\lambda_{\rm arc}$. %
                The rounder \gls{SIDM} population has a lower convexity rate, enabling a statistical test with many streams. %
                \autoref{sec:convexity_rate_grid_full} provides further details. %
                \label{fig:cosmology:convexity_rate_vs_flattening}
            }
        \end{figure*}

        % The needed rate
        In its most basic form the convexity test needs one number per model cosmology: the fraction of catalog streams in which we detect a convex segment. %
        The catalog contains streams with measured tracks; a stream counts as a detection if, given a \gls{CoM}, it shows a reliable convex segment (see \autoref{sec:streams:match_orbit}) that is long enough to recognize, which we quantify below. %
        Let $E$ be that detection event. %
        The population rate of model $\mcal{M}$ is the probability of $E$ given $\mcal{M}$, marginalized over the predicted halo-shape parameters $\mbf{S}$ and the viewing direction $\zhat$, %
        \begin{align}\label{eq:cosmology:p_M}
            \lamM
            &\equiv P(E\mid\mcal{M}) \\
            &= \int P(E\mid\mbf{S},\zhat)\,P(\mbf{S}\mid\mcal{M})\,P(\zhat)\,\dif{\mbf{S}}\,\dif\zhat, %
        \end{align}
        where $P(E\mid\mbf{S},\zhat)$ is the per-host convexity probability, $P(\mbf{S}\mid\mcal{M})$ the shape distribution predicted by $\mcal{M}$, and $P(\zhat)$ a (generally isotropic) viewing prior, taken independent of $\mbf{S}$. %
        We abbreviate these $p(\mbf{S},\zhat)$ and $P_{\mcal{M}}(\mbf{S})$. %
        We label the models $\mcal{M}\in\{\mrm{C},\mrm{SI}\}$DM, with convexity rates $\lamM$. %
        Because $\lamM$ averages over all of $P_{\mcal{M}}(\mbf{S})$, two models can predict overlapping shape distributions and still predict different convexity rates; we will see this explicitly in \autoref{fig:cosmology:convexity_rate_vs_flattening}. %

        % Computing the per-host probability
        The per-host probability $p(\mbf{S},\zhat)$ depends on model properties, especially halo shape and assembly history, which influence the distribution of stream progenitors. %
        Cosmological simulations or semi-analytic models can predict this distribution \citep[e.g.,][]{Dropulic+:2025}, which we need to test a specific dark matter model. %
        In \autoref{fig:cosmology:convexity_rate_vs_flattening}, we show $p(\mbf{S},\zhat)$ calculated through Monte-Carlo sampling of orbits and viewing angles in a Milky-Way-mass halo. %
        For each potential, we integrate \input{output/cosmology/n_orbits.tex} orbits for \input{output/cosmology/t_integrate.tex}, using apocentres and eccentricities drawn uniformly from \input{output/cosmology/r_range.tex} and $\input{output/cosmology/e_range.tex}$ respectively \citep{Sandford+:2017:QuantifyingTidalStream}, and initial velocity directions drawn uniformly over \input{output/cosmology/i_range.tex} in the local tangent plane. %
        Viewing directions are isotropic, with \input{output/cosmology/n_views.tex} external views.  %
        We discard projected segments within \input{output/cosmology/r_exclude.tex} of the \gls{CoM}, where galaxy light complicates ridge-line tracing. %
        Each host uses a triaxial \gls{NFW} density model (\texttt{galax.TriaxialNFWPotential}; \citealt{galax}), with potential calculated through the Chandrasekhar ellipsoidal formalism \citep{Chandrasekhar:1969:EllipsoidalFiguresEquilibrium, Binney+Tremaine:2008:GalacticDynamics}. %
        The density axis ratios are $b/a \equiv r_y/r_x$ and $c/a \equiv r_z/r_x$, with $r_x \ge r_y \ge r_z$. %
        The left panel of \autoref{fig:cosmology:convexity_rate_vs_flattening} shows $p(\mbf{S},\zhat)$ across a grid of shapes $\mbf{S}$ on the $(b/a,\,c/a)$ plane. %

        % The shape distribution
        The shape distribution $P_{\mcal{M}}(\mbf{S})$ is set by the dark matter model, and \cref{eq:cosmology:p_M} can use any supplied distribution from any model. %
        From cosmological simulations we take the \gls{CDM} distribution of $(b/a,\,c/a)$ \citep{Brinckmann+:2018:StructureAssemblyHistory,Giocoli+:2026:AIDATNGProject3D}, and convert each shape to a viewing-averaged projected flattening $\bar{q}$. %
        $\bar{q}$ is the minor-to-major axis ratio of the viewing-direction-averaged projected isodensity ellipse, and depends only on axis ratios $(b/a,\,c/a)$. %
        These regions describe halos at the radii traced by streams, beyond where baryons dominate the shape (\autoref{sec:cosmology:microphysics}). %
        Self-interaction rates fall with density, so \gls{SIDM} and \gls{CDM} halo shapes converge in the outskirts unless $\sigma/m$ is large \citep{Peter+:2013:CosmologicalSimulationsSelfInteracting}. %
        We therefore use the \gls{SIDM} region corresponding to $\sigma/m\gtrsim\text{a few}\ \unit{\square\centi\meter\per\gram}$, for which self-interactions round the halo well beyond the density core; the shape difference from \gls{CDM} exceeds the density-profile difference out to $r\approx0.25$--$0.4\,r_{\rm vir}$ \citep{Fischer+:2022:CosmologicalSimulationsRare,Fischer+:2024:CosmologicalIdealizedSimulations}. %

        % Non-monotonicity of b/a -> \bar{q}
        At fixed $c/a$, $\bar{q}$ is not monotonic in $b/a$. %
        It peaks near $b/a\!\sim\!\input{output/cosmology/qbar_turnover_ba_range.tex}$ and then decreases toward the oblate limit, $b/a=1$. %
        This decrease happens because the projected axis ratio along equatorial sightlines ranges from $c/a$ to $c/b$. %
        As $b/a\!\to\!1$, this range reduces to $c/a$. %
        For example, at $c/a=\input{output/cosmology/qbar_turnover_ca.tex}$, $\bar{q}$ decreases from $\input{output/cosmology/qbar_turnover_qbar_peak.tex}$ at $b/a=\input{output/cosmology/qbar_turnover_ba.tex}$ to $\input{output/cosmology/qbar_turnover_qbar_oblate.tex}$ at $b/a=1$; correspondingly, $\lamdadet{10}$ increases from $\input{output/cosmology/qbar_turnover_rate_peak.tex}\%$ to $\input{output/cosmology/qbar_turnover_rate_oblate.tex}\%$. %
        In the oblate limit, $\bar{q}=\int_{0}^{1}\sqrt{(c/a)^{2}+[1-(c/a)^{2}]\,(\cos{i})^{2}}\,\dif{\cos{i}}$, where $i$ is the inclination angle. %
        This expression agrees with our Monte-Carlo values to better than $10^{-4}$. %
        Thus, increasing $c/a$ always increases $\bar{q}$, but increasing $b/a$ decreases $\bar{q}$ beyond the peak. %
        The white contours in the left panel of \autoref{fig:cosmology:convexity_rate_vs_flattening} show this turnover near the oblate limit. %

        We evaluate the average over the full shape distribution $P_{\mcal{M}}(\mbf{S})$. %
        For \gls{CDM}, we use the Auriga halos described in \autoref{sec:cosmology:rates}, evaluating from each halo's shape $\bar{q}$ and convexity rate. %
        For \gls{SIDM}, we use the same halos after relaxation. %
        The two averages therefore compare similarly constructed populations. %
        Dot-dashed lines in the right panel of \autoref{fig:cosmology:convexity_rate_vs_flattening} mark the two population means; \autoref{fig:cosmology:convexity_rate_grid_full} adds shaded bands for their central $\input{output/cosmology/band_pctl.tex}\%$ in $\bar{q}$. %
        As a robustness check, we weight each shape region uniformly instead of using the actual distribution; this gives $\eta=\input{output/cosmology/eta_region_mean.tex}$ and changes the required $N$ by a factor of \input{output/cosmology/n_required_ratio.tex}, without changing the order-of-magnitude forecast in \autoref{sec:cosmology:sample_size}. %

        Any shape distribution, whether predicted theoretically or measured observationally, can enter \cref{eq:cosmology:p_M} without changing the framework. %
        Public \gls{SIDM} zoom-in suites now provide shape distributions for several \gls{SIDM} models \citep{Nadler+:2025:SIDMConcertoCompilation}. %
        These distributions could replace the relaxation model of \autoref{sec:cosmology:rates} with direct simulation results. %

        % What counts as a detection?
        Not every convex segment will be detectable in imaging. %
        Beyond the width and near-tangent conditions of \autoref{sec:streams:match_orbit}, a segment must be recognizably convex. %
        We investigate that requirement with three estimators, computed on the same orbits and views; survey-specifics such as surface brightness and track fitting we defer to \autoref{sec:discussion:calibration}. %
        One measure is the arc-length fraction $\lambda_{\rm arc}$, the convex fraction of the total projected track; it registers a detection only if the convex arcs together cover much of that track, however short or distributed they are. %
        Our second measure, $\lambda_{\rm det}^{(>\ell_{\min})}$, subdivides each \input{output/cosmology/t_integrate.tex} orbit into sliding \input{output/cosmology/window.tex} windows. %
        Each window better matches the lifetime of a dwarf-galaxy stream, and sliding the windows subsamples over the unknown initial orbital phase. %
        Additionally, $\lambda_{\rm det}^{(>\ell_{\min})}$ requires a convex segment longer than $\ell_{\min}$ for a detection. %
        We will take a conservative $\lamdadet{10}$ ($\ell_{\min} = \input{output/cosmology/ell_min.tex}$) so the convex arc must extend beyond dense baryonic structures like bars or disks. %
        The last measure is just the ideal detection rate $\lamdadet{0}$ at the $\ell_{\min}\!\to\!0$ limit, assuming convexity is measurable at any length. %
        We expect the real observable rate to lie between $\lambda_{\rm arc}$ and $\lamdadet{10}$. %
        For robustness, we adopt the more pessimistic $\lambda_{\rm arc}$ when quoting numbers. %

        % Reading the rate figure
        \autoref{fig:cosmology:convexity_rate_vs_flattening} shows $\lamdadet{10}$ and $\lambda_{\rm arc}$ across the grid of triaxial \gls{NFW} density models; \autoref{sec:convexity_rate_grid_full} adds $\lamdadet{0}$ and further details. %
        The left panel maps $\lamdadet{10}$ over $(b/a,\,c/a)$. %
        The rate decreases smoothly and nearly monotonically to zero as the density approaches the spherical limit, $(b/a,\,c/a)\to(1,1)$ (\autoref{sec:triaxial_potentials:axisymmetric}). %
        % What the two regions are
        The two regions in \autoref{fig:cosmology:convexity_rate_vs_flattening} show the joint interquartile regions of the \gls{CDM} and \gls{SIDM} halo populations in $(b/a,\,c/a)$: each rectangle contains half of its population. %
        For \gls{CDM}, we use the 30 Milky-Way-mass Auriga halos analyzed by \citet{Prada+:2019:DarkMatterHalo}, pooling their shapes over the radial range traced by streams. %
        Its region covers $b/a=\input{output/cosmology/cdm_ba_range.tex}$ and $c/a=\input{output/cosmology/cdm_ca_range.tex}$. %

        No existing cosmological hydrodynamical suite provides a statistical sample of galaxy-mass ($\qtyrange{e11}{e13}{\Msun}$) halos with \gls{SIDM} cross-sections large enough to round halos at the radii probed by streams, $\sigma/m\gtrsim\text{a few}\,\unit{\square\centi\meter\per\gram}$. %
        We therefore construct approximate \gls{SIDM} counterparts to the Auriga hydrodynamic \gls{CDM} halos by relaxing each axis ratio toward unity, with a scaling as follows. %
        Dark-matter-only simulations at high cross-section set the size of this shift, while lower-cross-section simulations with baryons calibrate the baryon-caused reduction of the halo sphericalization \citep{Giocoli+:2026:AIDATNGProject3D}. %
        Baryons similarly make \gls{CDM} and \gls{SIDM} halo shapes more alike in other simulations \citep{Despali+:2022:ConstrainingSIDMHalo,Vargya+:2022:ShapesMilkyWayMassGalaxies}. %
        The halos are relaxed by
        \begin{equation}\label{eq:cosmology:sidm_relaxation}
            s_{\mrm{SI}} = s_{\mrm{C}} + f\,(1 - s_{\mrm{C}}), \qquad s \in \{b/a,\ c/a\}, %
        \end{equation}
        with larger shift for less spherical halos. %
        We choose $f$ so that the median $c/a$, $\widetilde{c/a}=\input{output/cosmology/auriga_ca_median.tex}$, increases by $\delta$: $ f=\frac{\delta\,\widetilde{c/a}}{1-\widetilde{c/a}} =\input{output/cosmology/sidm_relaxation.tex}.$ %
        We take $\delta=\input{output/cosmology/sidm_ca_offset.tex}\%$ from \citet{Giocoli+:2026:AIDATNGProject3D}. %
        Their dark-matter-only halos show a $35.2\%$ offset, while baryons reduce this by a factor of $2.16$ over $\numrange{0.22}{0.37}\,R_{200}$, the radial range of $r_1$ for $\sigma/m\gtrsim\text{a few}$. %
        Thus $\delta=35.2\%/2.16$, leaving no free parameter. %
        \citet{Chua+:2021:ImpactInelasticSelfinteracting} further suggest that this offset saturates at these cross-sections. %
        This produces the high-$\sigma/m$ \gls{SIDM} distribution, whose region covers $b/a=\input{output/cosmology/sidm_ba_range.tex}$ and $c/a=\input{output/cosmology/sidm_ca_range.tex}$. %
        This extrapolation is adequate for forecasting how halo sphericalization changes the population convexity rate, but quantitative constraints on $\sigma/m$ require matched cosmological hydrodynamical \gls{CDM} and \gls{SIDM} ensembles in this mass, radial, and cross-section regime. %
        We label the mean projected flattenings of the two populations by $\bar{q}_{\mrm{C}}^{\rm sim}$ and $\bar{q}_{\mrm{SI}}^{\rm sim}$. %
        The forecast uses this population difference, quantified by the rate ratio $\eta$. %

        % Why this has to be a population measurement
        The right panel of \autoref{fig:cosmology:convexity_rate_vs_flattening} shows the same behavior for $\bar{q}$. %
        The envelopes' vertical spread between potentials at a common $\bar{q}$, is physical. %
        $\bar{q}$ is only the mean projected flattening, but convexity depends on the full intrinsic shape $S$. %
        At fixed $\bar{q}$ an oblate halo has a higher convexity rate than a prolate one. %
        The gray percentile bands show the orbit-to-orbit scatter. %
        That scatter is often larger than the separation between neighboring points. %
        Neither a single stream nor a single host can distinguish between models, but the population mean rate can. %

        % The calibrated rates
        Averaging $\lambda_{\rm arc}$ over each population (lower white diamonds) gives $\lamC\!\sim\!\input{output/cosmology/lambda_arc_cdm.tex}\%$ and $\lamSI\!\sim\!\input{output/cosmology/lambda_arc_sidm.tex}\%$. %
        With clean data and careful reduction the measured rates might approach $\lamdadet{10}$ (upper white diamonds), $\lamC\!\sim\!\input{output/cosmology/lambda_det_cdm.tex}\%$ and $\lamSI\!\sim\!\input{output/cosmology/lambda_det_sidm.tex}\%$. %
        The ratio of these two rates, $\eta=\input{output/cosmology/eta_fid.tex}$, determines the required catalog size to distinguish between models. %
        \autoref{tab:cosmology:lambdas} collects the convexity rates used in this section. %
        % Reference table for the many rates in play.
        \begin{table}[htbp]
            \centering
            \caption{Convexity rates used in \autoref{sec:cosmology}.}
            \label{tab:cosmology:lambdas}
            \footnotesize
            % The det-rate symbols carry a superscript containing a subscript, which
            % overflows the default row height and collides with the row below.
            \renewcommand{\arraystretch}{1.1}%
            \begin{tabular*}{\columnwidth}{@{\extracolsep{\fill}}ll}
                \hline\hline
                Symbol & Definition \\
                \hline
                % \multicolumn{2}{@{}l}{{Predicted:}} \\
                $\lamM$ & rate under model $\mcal{M}$ (\cref{eq:cosmology:p_M}) \\
                $\lamC$, $\lamSI$ & the $\mcal{M}=\text{\gls{CDM}}$ and $\mcal{M}=\text{\gls{SIDM}}$ rates \\
                $\eta$ & ratio $\lamSI/\lamC$ \\[2pt]
                % \multicolumn{2}{@{}l}{{Estimators of $\lamM$:}} \\
                $\lambda_{\rm arc}$ & convex arc-length fraction \\
                $\lamdadet{\ell_{\min}}$ & windows with a segment longer than $\ell_{\min}$ \\
                $\lamdadet{10}, \lamdadet{0}$ & the $\ell_{\min}\!=\!10$, and $\ell_{\min}\!\to\!0$ limit \\[2pt]
                % \multicolumn{2}{@{}l}{{Measured:}} \\
                $\lamObs$ & observed rate $K/N$ \\
                \hline
            \end{tabular*}
        \end{table}

    % sec:cosmology:rates (end)

    \subsection{Excluding a Model}\label{sec:cosmology:excluded}

        % Counting experiment
        We now ask whether the observed catalog can exclude either dark matter model. %
        We fix $\lamC$ and $\lamSI$ to the simulation-calibrated values from \autoref{sec:cosmology:rates}, leaving no free parameters in either model. %
        The catalog likelihood is then binomial, as in other population-proportion measurements based on counts \citep{Cameron:2011:EstimationConfidenceIntervals}. %
        Under model $\mcal{M}$,
        \begin{equation}\label{eq:cosmology:binom}
            K \mid \mcal{M} \sim \mathrm{Bin}(N,\lamM).
        \end{equation}

        % Rejecting a model
        Too many convexities rejects \gls{SIDM} while too few rejects \gls{CDM}. %
        Both are binomial tail probabilities, %
        \begin{align}
            p_{\mrm{SI}}^{+} &= P(K'\ge K\mid N,\lamSI),
                \quad
                K'\!\sim\!\mathrm{Bin}(N,\lamSI) \label{eq:cosmology:p_reject_sidm}
            \\
            p_{\mrm{C}}^{-} &= P(K'\le K\mid N,\lamC),
                \quad
                K'\sim\mathrm{Bin}(N,\lamC), \label{eq:cosmology:p_reject_cdm}
        \end{align}
        which we report as one-sided Gaussian significances $Z_{\mrm{rej},\mcal{M}} = \Phi^{-1}(1-p)$, with $\Phi$ the normal cumulative distribution. %

        % Exact tails
        For scaling arguments the normal approximation is convenient, %
        \begin{equation}\label{eq:cosmology:one_sided_rejection}
            Z_{\mrm{rej},\mcal{M}}
            \simeq \frac{\| \lamObs - \lamM \|}{\sqrt{\lamM (1 - \lamM)/N}}
        \end{equation}
        which shows the significance scaling with $N$ and with the difference between the observed and predicted rates. %
        We do not use the approximation for quoted values. %
        The approximation, called the Wald form, degrades at small $\lambda$ \citep{Brown+:2002:ConfidenceIntervalsBinomial,Cameron:2011:EstimationConfidenceIntervals}. %
        Larger $N$ improves but does not fix it. %
        For $\lamSI\!=\!\input{output/cosmology/lambda_arc_sidm.tex}\%$ it underestimates the required catalog by $\sim\!\input{output/cosmology/wald_error_1sigma.tex}\%$ at $1\sigma$. %
        We therefore use exact binomial tails for significances and the Wald form for forecasting, where it matches the exact calculation to within $\input{output/cosmology/wald_error_3sigma.tex}\%$ by $3\sigma$. %

        % Using the catalog
        Consider a catalog of $N=\input{output/cosmology/worked_n.tex}$ streams at the low rate estimate $\lambda_{\rm arc}$ of \autoref{sec:cosmology:rates}. %
        If the universe is \gls{CDM}, the median catalog contains $K=\input{output/cosmology/worked_k_cdm.tex}$ convexities, and \cref{eq:cosmology:p_reject_sidm} rejects \gls{SIDM} at $p_{\mrm{SI}}^{+}=\input{output/cosmology/worked_p_sidm.tex}$, or $Z_{\mrm{rej,SI}}=\input{output/cosmology/worked_z_sidm.tex}$. %
        If it is \gls{SIDM}, the median is $K=\input{output/cosmology/worked_k_sidm.tex}$, and \cref{eq:cosmology:p_reject_cdm} rejects \gls{CDM} at $p_{\mrm{C}}^{-}=\input{output/cosmology/worked_p_cdm.tex}$, or $Z_{\mrm{rej,C}}=\input{output/cosmology/worked_z_cdm.tex}$. %
        \Cref{eq:cosmology:one_sided_rejection} gives $\input{output/cosmology/worked_z_sidm_wald.tex}$ and $\input{output/cosmology/worked_z_cdm_wald.tex}$ for the same two catalogs, overstating the \gls{SIDM} rejection and understating the \gls{CDM} one. %
        All four are median expected significances, with half of all catalogs doing worse \citep{Cowan+:2011:AsymptoticFormulaeLikelihoodbased}. %
        A forecast needs more than the median outcome, and the sample-size calculation of \autoref{sec:cosmology:sample_size} will set a higher bar, estimating instead $N_\sigma=\input{output/cosmology/worked_n_sigma.tex}$. %

    % sec:cosmology:excluded (end)

    \subsection{Favoring a Model}\label{sec:cosmology:favored}

        \begin{figure}
            \centering
            \includegraphics[width=\columnwidth]{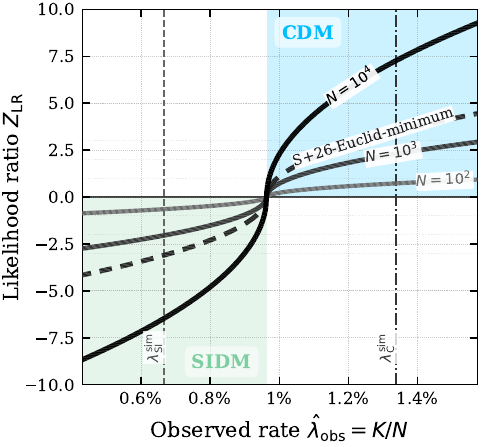}%
            \caption{%
                Likelihood-ratio $Z_{\rm LR}$ comparison of \gls{CDM} and \gls{SIDM} from the observed rate $\lamObs=K/N$, the $K$ streams containing a convex segment out of a catalog of $N$, for $N=\input{output/cosmology/catalog_sizes.tex}$ (solid). %
                The dashed curve is the $N\approx\input{output/cosmology/n_euclid_min.tex}$ streams projected for \euclid{} by \citet{Starkman+:2026:EuclidQ1}. %
                Curves use the predicted rates $(\lamC,\lamSI)\approx(\lamCval,\lamSIval)$. %
                $Z_{\rm LR}$ measures relative model preference: positive values favor \gls{CDM} (blue quadrant), negative values favor \gls{SIDM} (green quadrant), and larger absolute values mean stronger preference. %
                Shading marks the quadrant in which each model is favored; vertical lines mark the two predicted rates, from \autoref{fig:cosmology:convexity_rate_vs_flattening}. %
                % $Z_{\rm LR}$ relabels $\Delta\ln\mathcal{L}$ to fit a wide range on one axis; it is not a significance, since Wilks' theorem does not apply to these two models. %
                With $\sim\!10^3$ streams, a rate near either prediction begins to favor its corresponding model; $\sim\!10^4$ streams strongly distinguish between the two. %
                \label{fig:cosmology:zlr}
            }%
        \end{figure}

        The same likelihood also compares the two hypotheses directly. %
        When the hypotheses have no free parameters, the likelihood ratio is the mathematically optimal test \citep{Neyman+Pearson:1933:IXProblemMost}. %
        We write the log-likelihood ratio with positive values favoring \gls{CDM},
        \begin{align}\label{eq:cosmology:loglikelihood_ratio}
            \Delta\ln\mathcal{L}
            &\equiv
                \ln\frac{P(K\mid N,\lamC)}{P(K\mid N,\lamSI)} \\
            &=
                N\left[
                    \lamObs\ln\frac{\lamC}{\lamSI}
                    +
                    (1-\lamObs)\ln\frac{1-\lamC}{1-\lamSI}
                \right].  \nonumber %
        \end{align}
        At fixed observed rate the log-likelihood ratio scales linearly with $N$. %
        With a flat prior over $\{\mrm{C},\mrm{SI}\}$DM, the posterior odds in favor of \gls{CDM} are $\exp(\Delta\ln\mathcal{L})$, the Bayes factor for the pair \citep{Kass+Raftery:1995:BayesFactors,Trotta:2008:BayesSkyBayesian}. %
        Both hypotheses are simple, so the Bayes factor is the likelihood ratio itself, with no evidence integral to evaluate. %
        In \autoref{fig:cosmology:zlr}, we use the signed likelihood-ratio scale \citep{Cowan+:2011:AsymptoticFormulaeLikelihoodbased}
        \begin{equation}\label{eq:cosmology:Z_LR}
            Z_{\mrm{LR}}
            =
            \mathrm{sign}(\Delta\ln\mathcal{L})
            \sqrt{2|\Delta\ln\mathcal{L}|}. %
        \end{equation}
        $Z_{\mrm{LR}}$ relabels $\Delta\ln\mathcal{L}$ and is a likelihood ratio, not a significance. %
        $\sqrt{2\Delta\ln\mathcal{L}}$ is the usual factor mapping a likelihood ratio to $N\sigma$, but that conversion needs Wilks' theorem \citep{Wilks:1938:LargeSampleDistributionLikelihood}, which does not apply here: \gls{CDM} and \gls{SIDM} are not nested models, and both rates are fixed a priori \citep{Protassov+:2002:StatisticsHandleCare,Algeri+:2020:SearchingNewPhenomena}. %
        So we use $Z_{\mrm{LR}}$ only to fit a wide range of $\Delta\ln\mathcal{L}$ on one axis, and read the model preference from the actual odds $\exp(\Delta\ln\mathcal{L})$. %

        % Reading the figure
        \autoref{fig:cosmology:zlr} shows how to use a catalog to constrain dark matter. %
        For a fixed catalog size, an observed rate near $\lamSI$ gives negative $Z_{\mrm{LR}}$ (\cref{eq:cosmology:Z_LR}) and favors \gls{SIDM}, while an observed rate near $\lamC$ gives positive $Z_{\mrm{LR}}$ and favors \gls{CDM}. %
        Increasing $N$ steepens the curve in \autoref{fig:cosmology:zlr} because the same observed rate difference carries more binomial information. %
        The horizontal axis is the actual measurement, $K/N$, so we can read the figure directly once we have a catalog. %

    % sec:cosmology:favored (end)

    \subsection{Distinguishing Dark Matter with Stream Catalogs}\label{sec:cosmology:sample_size}

        \begin{figure}
            \centering
            \includegraphics[width=\columnwidth]{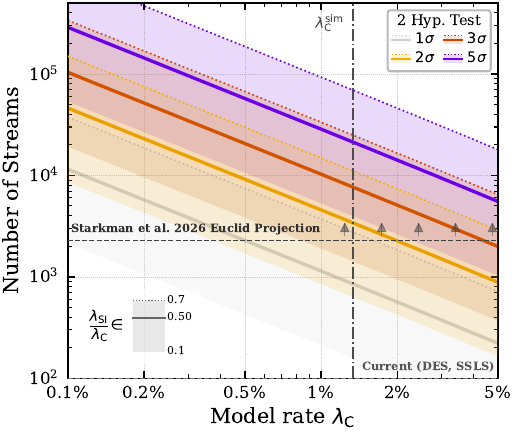}%
            \caption{%
                Required stream sample size $N$ to distinguish \gls{CDM} from \gls{SIDM} to some $\sigma$ (\cref{eq:cosmology:N_required_full}), as a function of the \gls{CDM} convexity rate $\lamC$. %
                Solid lines assume that, as suggested by our simulations, \gls{SIDM} produces $\eta\equiv\lamSI/\lamC=\input{output/cosmology/eta_fid.tex}$ times the \gls{CDM} rate of convex streams; shaded bands show the simulated range $\eta\in[\input{output/cosmology/eta_lo.tex},\input{output/cosmology/eta_hi.tex}]$. %
                A larger $\lamC$ or a smaller $\eta$ (stronger \gls{SIDM} sphericalization) both reduce the required sample size. %
                Horizontal dashed line marks the very lower bound of expected stream-catalog sizes for \euclid{} (see \!\!\citet{Starkman+:2026:EuclidQ1}), with more realistic predictions, and including \gls{Rubin}/\gls{Roman}, expecting an order-of-magnitude larger sample. %
                A vertical dot-dashed line marks the \gls{CDM} convexity rate, $\lamC^{\mrm{sim}}$, predicted by our illustrative simulation (\autoref{sec:cosmology:rates}). %
                This is a forecast. %
                It neglects observer effects and the cosmological evolution of halo shapes; both modulate the effective $\lamC$ and $\eta$. %
                For the simulated convexity rates shown here, large catalogs of streams can distinguish between \gls{CDM} and \gls{SIDM}. %
                \label{fig:cosmology:Nreq}
            }%
        \end{figure}

        The likelihood of \cref{eq:cosmology:binom} fixes the catalog size needed to distinguish the two models at a given significance. %
        We measure that separation through the convexity-rate ratio $\eta\equiv \lamSI/\lamC$, with $\eta\lesssim 1$ since \gls{SIDM} generally predicts a more spherical halo population. %

        For a one-sided separation at $N_\sigma$, with significance and power set equal ($N_{\sigma,\alpha}=N_{\sigma,\beta}=N_\sigma$), the required sample size in the normal approximation is \citep[e.g.,][]{Fleiss+:2003:StatisticalMethodsRates}
        \begin{align}
            N
            &\ge
                N_\sigma^2\!\left(\tfrac{\sqrt{\lamC(1-\lamC)}+\sqrt{\lamSI(1-\lamSI)}}{\lamC-\lamSI}\right)^{\!2}
                \label{eq:cosmology:N_required_full} \\
            &\gtrsim
                \frac{N_\sigma^2}{\lamC\,(1-\sqrt{\eta})^{2}},
                \qquad \lamC\ll 1 \ (100\%),
                \label{eq:cosmology:N_required} %
        \end{align}
        the second line using $\sqrt{\lambda(1-\lambda)}\simeq\sqrt{\lambda}$. %
        The required $N$ scales inversely with both $\lamC$ and $(1-\sqrt{\eta})^{2}$. %
        Taking the low rate estimate of \autoref{sec:cosmology:rates}, \cref{eq:cosmology:N_required_full} gives $N\!\sim\!\input{output/cosmology/n_required_fiducial.tex}$ for $N_\sigma=1,2,3,5$. %
        The optimistic $\lamdadet{10}$ calibration requires a much smaller catalog, $N\!\sim\!\input{output/cosmology/n_required_optimistic.tex}$ for the same significances. %
        The suppression ratio is nearly unchanged ($\eta=\input{output/cosmology/eta_det.tex}$), so a rate \input{output/cosmology/lambda_det_over_fiducial.tex} times larger reduces the required $N$ by close to the same factor, and a $3\sigma$ test needs only $\sim\!\input{output/cosmology/n_required_optimistic_3sigma.tex}$ streams. %
        \Cref{fig:cosmology:Nreq} plots \cref{eq:cosmology:N_required_full} over $(\lamC,\eta,N_\sigma)$. %
        The horizontal axis shows $\lamC$, calibrated in \autoref{sec:cosmology:rates}, while $\eta$ sets the band width. %
        The dependence on $\eta$ is steep. %
        At the value $\lamC\!\sim\!\lamCval$ motivated by \autoref{fig:cosmology:convexity_rate_vs_flattening}, moving across the band from $\eta=\input{output/cosmology/eta_lo.tex}$ to $\eta=\input{output/cosmology/eta_hi.tex}$ raises the $3\sigma$ requirement from $N\!\sim\!\input{output/cosmology/n_required_3sigma_eta_lo.tex}$ to $N\!\sim\!\input{output/cosmology/n_required_3sigma_eta_hi.tex}$. %
        The curves in \cref{fig:cosmology:Nreq} are forecasts, not the significance of an observed catalog. %
        For a measured $K=N\lamObs$, \cref{eq:cosmology:p_reject_sidm,eq:cosmology:p_reject_cdm,eq:cosmology:loglikelihood_ratio} give the model rejection and model comparison directly, as worked through in \autoref{sec:cosmology:excluded}. %

        % Reading the figure
        The vertical band-width in \cref{fig:cosmology:Nreq} reflects uncertainty in the shape-distribution prediction $P_{\mcal{M}}(\mbf{S})$, which fixes $\eta$. %
        The horizontal position of the dot-dashed line reflects uncertainty in the per-host convexity probability $p(\mbf{S},\zhat)$, which fixes $\lamC$. %
        Reading at the point $(\lamC,\eta)\!\sim\!(\lamCval,\input{output/cosmology/eta_fid.tex})$, current Local-Group catalogs ($N\!\sim\!\numrange{30}{100}$) remain well below $1\sigma$, and even $N\!\sim\!100$ corresponds to only $\sim\!\input{output/cosmology/n_sigma_local_group.tex}\sigma$. %
        We are currently sample-limited. %
        A $3\sigma$ separation between \gls{CDM} and \gls{SIDM} is plausible in the \euclid{}/\gls{Rubin}/\gls{Roman} era, and $5\sigma$ needs the largest \gls{Roman}-scale catalogs. %

        % Two simplifications in the figure
        \Cref{fig:cosmology:Nreq} holds $\lamC$ and $\eta$ fixed, but neither is actually a cosmological constant. %
        The halo-shape distribution $P_{\mcal{M}}(\mbf{S})$ evolves with redshift, depends on host mass, and for \gls{SIDM} depends on the cross-section $\sigma/m$ \citep{Vera-Ciro+:2011:ShapeDarkMatter,Vargya+:2022:ShapesMilkyWayMassGalaxies,Bautista+:2025:JeansModelShapes}. %
        Hosts are more triaxial at high $z$ and round over cosmic time as adiabatic contraction and elastic scatterings act. %
        A catalog therefore samples many points on this plane rather than one, and its forecast is a survey-weighted average over redshift and host-mass bins. %
        A high-$z$ subsample probes a different point on the $(\lamC,\eta)$ plane than a $z\!\sim\!0$ subsample. %
        The curves assume the stream its $\|(\xsky^{\pm})_{\perp}\|\gg W$ are well measured, so they are lower bounds for an unselected sample. %
        The survey selection function and the orbital prior both enter $p(\mbf{S},\zhat)$, shifting the effective $(\lamC,\eta)$ away from the simulation calibration and so shifting the curves relative to the survey lines. %
        We discuss this calibration in \autoref{sec:discussion:calibration}. %
        Where the forecast falls within that significance band depends on the \gls{SIDM} microphysics, since the cross-section sets $\eta$. %

    % sec:cosmology:sample_size (end)

% sec:cosmology (end)

\section{Discussion}\label{sec:discussion}

    % Intro
    The convexity criterion of \autoref{sec:diagnostic} works at several scales. %
    For a single stream, from the projected track and \gls{CoM} a resolved convex segment rules out the whole \gls{LoS}-axisymmetric centrally-concentrated family of potentials, requiring neither kinematics nor distances (\autoref{sec:triaxial_potentials:axisymmetric}). %
    For a single host galaxy, the fraction of its streams with a convex segment is the per-host probability $p(\mbf{S},\zhat)$, evaluated at that host's shape $\mbf{S}$ and \gls{LoS} $\zhat$ (\autoref{sec:cosmology:rates}). %
    Across a population, averaging $p(\mbf{S},\zhat)$ over a cosmological shape distribution gives a rate $\lamM$ that differs between the \gls{CDM} and \gls{SIDM} populations considered here. %
    A stream catalog can therefore test these population-level predictions. %
    In \autoref{sec:cosmology}, we forecast the sample size needed to distinguish them. %
    % Roadmap for whole discussion.
    Here we discuss the connection between this forecast and dark-matter microphysics, including the calibrations required to turn a measured convexity rate into constraints on the self-interaction cross-section (\autoref{sec:cosmology:microphysics}). %

    \subsection{Convexity Rates Constrain the SIDM Cross-Section}\label{sec:cosmology:microphysics}

        % Roadmap
        We now connect the rate forecast of \autoref{sec:cosmology} to the \gls{SIDM} self-interaction cross-section. %
        The convexity test directly measures the rate ratio $\eta\equiv\lamSI/\lamC$, not the cross-section. %
        To convert between them, simulations must predict the \gls{SIDM} halo-shape distribution $P_{\mrm{SI}}(\mbf{S})$ as a function of interaction strength. %
        $\sigma/m$ parametrizes this strength for velocity-independent models; more generally the velocity-averaged interaction rate $\langle\sigma v\rangle/m$ sets it. % 
        This distribution sets $\lamSI$ through \cref{eq:cosmology:p_M}, and therefore sets $\eta$ once the comparator rate $\lamC$ is fixed \citep[for reviews see][]{Tulin+Yu:2018:DarkMatterSelfinteractions,Adhikari+:2025:AstrophysicalTestsDark}. %
        Comparing a measured $\lamObs$ with these predicted rates then constrains the range of $\sigma/m$ allowed by the stream catalog. %

        % Relating $\eta$ to $\sigma/m$.
        Self-interactions redistribute heat and isotropize the velocity tensor on a relaxation time $t_{\mrm{rel}}\sim[\rho\,\langle\sigma v\rangle/m]^{-1}$ \citep{Spergel+Steinhardt:2000:ObservationalEvidenceSelfInteracting}. %
        A halo becomes rounder where $t_{\mrm{rel}}$ is shorter than its age and approaches \gls{CDM}-like triaxiality elsewhere, like at large radii \citep{Dave+:2001:HaloPropertiesCosmological,Peter+:2013:CosmologicalSimulationsSelfInteracting}. %
        In the spherical limit the per-host convexity probability $p(\mbf{S},\zhat)$ goes to zero (\autoref{sec:diagnostic}). %
        Increasing $\sigma/m$ therefore rounds a larger part of the halo, suppresses $\lamSI$, and lowers $\eta$. %
        A lower $\eta$ separates the two model rates more strongly and reduces the number of streams required at fixed significance (\autoref{sec:cosmology:sample_size}). %

        % Radius of sensitivity 
        The mapping from $\eta$ to $\sigma/m$ also depends on the radii probed by the streams. %
        We exclude the inner halo, where baryons dominate the mass budget. %
        At large radii, observable streams become increasingly rare. %
        For galaxies, with $M(r<R_{200})\sim\qty{e12}{\Msun}$ and $R_s\approx\qtyrange{15}{20}{\kilo\parsec}$, we focus on $\qty{30}{\kilo\parsec}\lesssim r\lesssim\qty{130}{\kilo\parsec}$, the range over which we measure the halo-shape distributions in \autoref{sec:cosmology:rates}. %
        The hydrodynamic calibration includes the effect of baryons on halo shape. %
        Hydrodynamic simulations over a wider range of \gls{SIDM} cross sections would further improve this calibration. %

        % Radius of sphericalization
        Within this range, the test is informative only where \gls{SIDM} and \gls{CDM} predict different shapes. %
        In simulations of galaxies, $\sigma/m\sim \qty{1}{\square\centi\meter\per\gram}$ rounds the inner halo, whereas $\sigma/m\sim\qty{10}{\square\centi\meter\per\gram}$ can extend sphericalization beyond the scale radius \citep{Peter+:2013:CosmologicalSimulationsSelfInteracting,Chua+:2021:ImpactInelasticSelfinteracting,Fischer+:2024:CosmologicalIdealizedSimulations,Bautista+:2025:JeansModelShapes}. %
        Cluster-scale and early-type simulations show the same qualitative behavior \citep{Brinckmann+:2018:StructureAssemblyHistory,Robertson+:2019:ObservableTestsSelfinteracting,Despali+:2022:ConstrainingSIDMHalo}. %
        Streams around disk and early-type galaxies should therefore be constraining for \gls{SIDM} models where $\sigma/m\gtrsim\text{a few}\ \unit{\square\centi\meter\per\gram}$, for which the shape difference to \gls{CDM} extends beyond the inner galaxy. %

        % Disks
        Baryons can round a halo at the same radii. %
        In disk galaxies, adiabatic contraction sphericalizes the halo within $r\!\lesssim\!0.1\,R_{200}\!\approx\!\qty{20}{\kilo\parsec}$ \citep{Kazantzidis+:2004:EffectGasCooling,Sameie+:2018:ImpactBaryonicDiscs,Chua+:2022:ImpactGalacticFeedback}. %
        This contracted region overlaps the region affected by \gls{SIDM} with $\sigma/m\sim\qty{1}{\square\centi\meter\per\gram}$ \citep{Peter+:2013:CosmologicalSimulationsSelfInteracting,Brinckmann+:2018:StructureAssemblyHistory}, so contraction and self-interactions can produce similar shape changes in the inner halo. %
        The severity of the degeneracy depends on how far out self-interactions reshape the halo, and simulations do not yet agree. %
        Differences between \gls{SIDM} and \gls{CDM} persist farther out in the halo axis ratios than in the spherically averaged density profiles, reaching $r\approx0.25$--$0.4\,r_{\rm vir}$ \citep{Fischer+:2022:CosmologicalSimulationsRare}. %
        For example, \citet{Fischer+:2024:CosmologicalIdealizedSimulations} find that a Milky-Way-mass halo with $\sigma/m=\qty{1}{\square\centi\meter\per\gram}$ remains rounder than its \gls{CDM} counterpart out to $\sim\!0.5\,R_{\rm vir}$ in dark-matter-only simulations, beyond which the halo-to-halo scatter exceeds the difference. %
        Stream convexities may therefore remain sensitive to this cross-section. %
        We nevertheless use the more conservative regime in which sphericalization extends beyond the scale radius \citep{Vera-Ciro+:2011:ShapeDarkMatter}, corresponding to $\sigma/m\gtrsim\text{a few}\ \unit{\square\centi\meter\per\gram}$ and approximately the $\eta\lesssim\input{output/cosmology/eta_sidm_regime.tex}$ part of \cref{fig:cosmology:Nreq}. %

        % Elliptical galaxies
        Elliptical hosts may extend the test to weaker self-interactions. %
        In the absence of a disk, their \gls{CDM} halos remain triaxial down to smaller radii \citep{Vera-Ciro+:2011:ShapeDarkMatter,Despali+:2022:ConstrainingSIDMHalo,Bautista+:2025:JeansModelShapes}. %
        The usable radial range can therefore extend inward toward $R_s$, where cross-sections below $\qty{1}{\square\centi\meter\per\gram}$ may still alter the halo shape, though halo shapes lose their constraining power as $\sigma/m$ falls toward $\qty{0.1}{\square\centi\meter\per\gram}$ \citep{Peter+:2013:CosmologicalSimulationsSelfInteracting}. %
        Baryons suppress the difference between \gls{CDM} and \gls{SIDM} halo shapes at the smaller radii probed by strong lensing and X-ray observations \citep{Despali+:2022:ConstrainingSIDMHalo}, but stream convexities probe farther out, at \input{output/cosmology/r_range.tex}. %
        Together, disk and elliptical hosts probe $\sigma/m\gtrsim\qty{1}{\square\centi\meter\per\gram}$, and possibly $\sigma/m\sim\qtyrange{0.1}{1}{\square\centi\meter\per\gram}$ if baryons do not erase the halo-shape difference in elliptical hosts. %

        % Mapping to cross-section
        For galactic halos, current simulations suggest that the range shown in \cref{fig:cosmology:Nreq}, $\eta\in[\input{output/cosmology/eta_lo.tex},\input{output/cosmology/eta_hi.tex}]$, corresponds roughly to $\sigma/m\sim\qtyrange{1}{10}{\square\centi\meter\per\gram}$ \citep{Vargya+:2022:ShapesMilkyWayMassGalaxies,Brinckmann+:2018:StructureAssemblyHistory,Fischer+:2024:CosmologicalIdealizedSimulations,Sameie+:2021:CentralDensitiesMilky}. %
        We therefore interpret our forecasts over this $\eta$ range as sensitivity to cross-sections of this order. %

        Stream convexities probe host halos, whose velocity dispersions, $\langle v\rangle\sim\qty{200}{\kilo\meter\per\second}$, are well below the cluster-scale dispersions, $\langle v\rangle\sim\qty{1000}{\kilo\meter\per\second}$, that set many existing constraints on $\sigma/m$. %
        If the \gls{SIDM} cross-section depends on velocity, stream convexities therefore probe an underexplored interaction regime \citep{Tulin+Yu:2018:DarkMatterSelfinteractions,Adhikari+:2025:AstrophysicalTestsDark}. %

    % sec:cosmology:microphysics (end)

    \subsection{Scope and Assumptions}\label{sec:discussion:assumptions}

        Throughout this paper, we make no assumptions about the progenitor, but we do assume the following about the host potential and the stream: %

        % \subsubsection{Host Potential}\label{sec:discussion:assumptions:host}

        \emph{Single-component effective potential.} %
        A stream may pass near a disk or bar at pericenter, but its convexity at a given location is determined mostly by the local potential rather than by its full orbital history. %
        We therefore restrict the diagnostic to convexity events in the outer halo, where the dark-matter triaxiality signal dominates. %
        For galactic halos, this is roughly $\qty{30}{\kilo\parsec} \lesssim r \lesssim \qty{200}{\kilo\parsec}$ (\autoref{sec:cosmology:microphysics}). %

        \emph{Time independence and satellite perturbations.} %
        The diagnostic assumes that the host potential is approximately static. %
        Massive satellites can violate this assumption by perturbing the stream directly or deforming the host halo, like the \gls{LMC} does in the Milky Way \citep{Erkal+:2019:TotalMassLarge,Shipp+:2021:MeasuringMassLarge,GaravitoCamargo+:2021:QuantifyingImpactLMC,Lilleengen+:2022:EffectDeformingDark,Brooks+:2024:LMCCallsMilkyWay}. %
        Such perturbations can create stream convexities in addition to the halo-shape signal. %
        This is consistent with the FIRE-2 forward models of \citet{Pearson+:2026:ExtragalacticStellarStreams}: their long, thin Curlicue stream is recovered without bias in a static potential, whereas the width-varying Massive stream yields a biased halo fit and the S-shaped stream provides no usable constraint outside its progenitor. %
        Therefore, the morphological criteria in \autoref{sec:streams:match_orbit} help identify streams for which a static-potential convexity analysis is appropriate. %
        Nearby satellites or visible stream features can flag perturbed systems \citep{Dillamore+:2022:ImpactMassiveSagittarius,Foote+:2025:Segue2Recently,Bonaca+:2020:VariationsWidthDensity}. %
        Nevertheless, isolated streams are the cleanest targets. %
        Future applications should exclude or downweight systems with nearby massive satellites or visible signs of stream perturbations. %
        \autoref{sec:discussion:outlook} discusses their identification in mock extragalactic images. %

        \emph{A known \gls{CoM}.} %
        The test requires the projected \gls{CoM} of the host because this point defines the radial direction, which in \cref{eq:theory:path:alignment_def} determines whether the stream contains convex segments. %
        We take the photometric center as an estimate for the \gls{CoM}. %
        Any uncertainty or offset between the two can be straightforwardly included by marginalizing $\lamM$ over the possible center positions (\autoref{sec:diagnostic}). %

        % \subsubsection{Stream Model}\label{sec:discussion:assumptions:stream}

        \emph{Stream coherence.} %
        We derive the convexity criterion of \autoref{sec:diagnostic} for an orbit but apply it on a stream ridge line. %
        As shown in \autoref{sec:streams:match_orbit}, the ridge line preserves the local concavity of the underlying orbits when the stream is coherent and its width varies smoothly. %
        These conditions can fail in many places, such as near an intact progenitor, in shell-forming debris, or where the stream fans. %
        Such features are generally recognizable in imaging. %
        A recently dissolved progenitor, however, may leave only a pair of tails. %
        A convexity near the junction of two such tails should similarly be excluded because it comes from the progenitor, not the host potential. %
        Simulated particle-spray and $N$-body streams support these results (\autoref{sec:triaxial_potentials:simulations}).

    % sec:discussion:assumptions (end)

    \subsection{Related Works}\label{sec:discussion:related_works}

        The closest comparable method is the curvature method of \citetalias{Nibauer+:2023}. %
        \citetalias{Nibauer+:2023} compare the observed curvature to a projected acceleration field from a trial potential. They vary over halo shapes to determine which produces agreement between the stream's track and underlying potential. %
        Our test compares the stream's track to an analytic reference direction ($\xcom$) supplied by the model class being falsified (here, the \gls{LoS}-axisymmetric centrally-concentrated family, \autoref{sec:triaxial_potentials:axisymmetric}). %
        Our diagnostic is a limiting case of \citetalias{Nibauer+:2023}, in which their fitted acceleration becomes analytic and their pointwise likelihood reduces to a binary falsification test. %
        This simplification means we can constrain the maximum flattening, but not the minimum flattening. %
        In trade, our approach needs only one input per stream: whether a convexity exists. %

        Convex segments are the interesting case. %
        In Figure 5 of \citetalias{Nibauer+:2023}, stream C contains a convexity and their method correctly rules out a spherical potential in favor of a prolate one. %
        Our approach rules out a spherical potential as well as all \gls{LoS}-axisymmetric centrally-concentrated potentials simply from the existence of the convex segment. %
        The two tests therefore answer different questions: \citetalias{Nibauer+:2023} fit a potential to recover its parameters, while we instead falsify a potential family without fitting any specific potential. %

        \citet{Chemaly+:2026:HierarchicalBayesianInference} pursue a complementary route through the same stream-shape data. %
        Their hierarchical Bayesian inference returns the full per-host shape posterior, using informative priors and a parametric halo-shape model to extract information from each stream. %
        The method developed here is a much coarser measurement, and it needs significantly larger catalogs for the same constraining power. %
        In trading specificity for volume, our method needs no detailed modeling of individual systems: no galaxy potential model, no careful characterization of each stream beyond whether it has a convexity, and no priors on the progenitor. %

        \citet{Bariego-Quintana:2024:TorsionStellarStreams} develop a complementary three-dimensional approach for Milky Way streams. %
        Rather than projected curvature, they use the torsion $\torsion$ of streams. %
        In a spherical potential orbits are planar and $\torsion = 0$. %
        A prolate or cylindrical halo gives a non-zero torsion that grows with the elongation. %
        Applied to Galactic streams, they find torsions consistent with a non-spherical Milky Way halo. %
        The torsion diagnostic requires the full $(x, y, z)$ track and distance information along the stream, so it is applicable to the Milky Way. %
        \citeauthor{Bariego-Quintana:2024:TorsionStellarStreams} acknowledge this directly, noting that at extragalactic distances only the projected curvature is measurable, not the torsion. %
        Our test is a projected counterpart of theirs, and is applicable across large extragalactic samples. %
        % We give up the out-of-plane information in $\torsion$ to gain a diagnostic applicable statistically across large extragalactic samples, without distance measurements or kinematics. %

        The \gls{Sgr} stream provides two examples of geometric constraints on the Galactic potential. %
        \citet{Johnston+:2005:TwoMicronAllSkySurvey} measured orbital-plane precession from the offset between the orbital poles of the leading and trailing tails, constraining the halo flattening. %
        \citet{Belokurov+:2014:PrecessionSagittariusStream} measured apsidal precession from the two tails' apocentres, constraining the radial density profile. %
        Both analyses need distances along the stream, whereas our test uses only the projected track. %

    % sec:discussion:related_works (end)

    \subsection{Calibrating the Population Rate}\label{sec:discussion:calibration}

        For the binomial likelihood of \cref{eq:cosmology:binom}, we use the simulation-calibrated detection rate $\lambda_{\rm det}$: the probability that an observed stream contains at least one detectable convex segment. %
        \Cref{fig:cosmology:convexity_rate_vs_flattening,fig:cosmology:convexity_rate_grid_full} show this rate alongside the ideal limit $\lamdadet{0}$ and the more conservative arc-length rate $\lambda_{\rm arc}$ (\autoref{sec:cosmology:rates}). %
        The halo-shape distributions, the per-host convexity probability, and the selection function are three ingredients of \cref{eq:cosmology:p_M} that require calibration before \cref{eq:cosmology:N_required} gives a quantitative bound. %
        \Cref{fig:cosmology:convexity_rate_vs_flattening,fig:cosmology:Nreq} use simple approximations sufficient for the order-of-magnitude forecast here. %

        \emph{The halo-shape distributions.} %
            $P_{\mrm{C}}(\mbf{S})$ and $P_{\mrm{SI}}(\mbf{S})$ depend on mass, redshift, and the assumed \gls{SIDM} cross-section $\sigma/m$, and vary with radius even at $z=0$ \citep{Vera-Ciro+:2011:ShapeDarkMatter}. %
            The relevant mass range is $M_{200}\!\sim\!\qtyrange{e12}{e13}{\Msun}$. %
            In this regime, shape differences persist between \gls{CDM} and \gls{SIDM} even where the spherically averaged profiles agree \citep{Bautista+:2025:JeansModelShapes,Giocoli+:2026:AIDATNGProject3D}. %
            In \cref{fig:cosmology:convexity_rate_vs_flattening}, we fix a measured \gls{CDM} shape distribution and a relaxed \gls{SIDM} counterpart. %
            This is a reasonable $z\!\sim\!0$ approximation for galaxy-mass halos, but self-interactions weaken with radius, so a precision analysis will require radially resolved shape distributions and stream samples from a new \gls{SIDM} simulation suite. %

        \emph{The per-host convexity probability $p(\mbf{S},\zhat)$.} %
            The heat map in \cref{fig:cosmology:convexity_rate_vs_flattening} Monte Carlo samples representative thin-stream orbits in each triaxial \gls{NFW} density model, with a Local-Group-inspired orbital prior and isotropic viewing directions. %
            It assumes an \gls{NFW} host without baryons and does not yet vary the orbital prior across the simulation grid. %
            The gray percentile bands show the resulting orbit-to-orbit scatter. %

            We can use the observed stream GD-1 (\autoref{fig:gd1_convexity}) as a qualitative check on this Monte Carlo approach. %
            GD-1 contains a convex segment longer than $\ell_{\min}=\input{output/cosmology/ell_min.tex}$ in \input{output/gd1/frac_det_ell_min.tex}\% of \input{output/gd1/n_views.tex} isotropic viewing directions. %
            That fraction lies within the Monte-Carlo predicted range for the different $\lamM$ measures, though GD-1 is not drawn from our static host models or orbital prior, so this is a qualitative check only. %

        \emph{The selection function.} %
            Both rates, $\lamC$ and $\lamSI$, assume all streams are observable, have clear tracks, and that their convex segments lie outside the near-tangent strip ($\|(\xsky^{\pm})_{\perp}\|\gg W$). %
            The selection function for these conditions varies based on the survey, catalog, and stream-detection pipeline (\autoref{sec:intro}). %
            The survey-aware statistic measures the rate of observing at least one convexity given that selection function.  %
            Computing it requires forward-modeling each candidate catalog (\gaia, \rubinlsst{}, \euclid{}), which we leave to future work. %
            Our three estimators differ by factors like spatial coverage, track recovery, smoothing, and detection thresholds, modeled through the minimum projected length $\lamdadet{\ell_{\min}}$. %
            \Cref{fig:cosmology:Nreq} uses the fiducial detection rate $\lambda_{\rm det}(\input{output/cosmology/ell_min.tex})$ from \cref{fig:cosmology:convexity_rate_vs_flattening}, which applies a minimal selection: a projected length cut and the \input{output/cosmology/r_exclude.tex} \gls{CoM} exclusion.  %
            $\lambda_{\rm arc}$ and $\lamdadet{0}$ represent the plausible range for a given survey. %
            A survey approaches $\lamdadet{0}$ when short convex segments resolve along the visible track, and $\lambda_{\rm arc}$ when the selection function down-weights the configurations that produce convexity. %
            For full inference, the selection function must be modeled or matched between simulations and observations. %

        Once we calibrate these ingredients, \cref{eq:cosmology:loglikelihood_ratio} gives the model comparison and \cref{eq:cosmology:N_required} gives the sample-size requirement. %
        Cosmological evolution of $P_{\mcal{M}}(\mbf{S})$ implies that a heterogeneous survey requires a weighted integral over redshift- and mass-binned versions of \cref{fig:cosmology:Nreq}, with $\lamM$ in \cref{eq:cosmology:N_required_full} matched to the observed redshift--mass distribution. %

    % sec:discussion:calibration (end)

    \subsection{Observational Outlook}\label{sec:discussion:outlook}

        We designed the diagnostic for surveys that resolve stream morphology but have no kinematics. %
        \euclid{} \citep{EuclidCollaboration+:2025:EuclidQuickData,Euclid+Walmsley+:2025:Q1} resolves extragalactic streams over a wide field, \rubinlsst{} \citep{Ivezic+:2019:LSSTScienceDrivers} extends Local-Group depth, and \gls{Roman} \citep{Spergel+:2015:WideFieldInfrarRedSurvey,Dey+:2023:RomAndromedaRoman} reaches high source densities at small angular scales. %
        None of these surveys delivers proper motions for the extragalactic populations they resolve, so six-dimensional phase space is unavailable for essentially every stream beyond the Local Group, and with it the 3D torsion of \citet{Bariego-Quintana:2024:TorsionStellarStreams}. %
        The curvature diagnostic requires only the projected track and projected \gls{CoM}. %

        The stream catalogs needed for this test are already being built. %
        Wide-field imaging searches of nearby galaxies have identified roughly one hundred dwarf stellar streams \citep{Martinez-Delgado+:2010:StellarTidalStreams,MartinezDelgado+:2023:HiddenDepthsLocal,Martinez-Delgado+:2025:StellarTidalStreams,Miro-Carretero+:2024:ExtragalacticStellarTidal,Miro-Carretero+:2025:ExtragalacticStellarTidal,Sola+:2025}. %
        Detailed modeling of individual streams has already constrained their host-halo shapes \citep{Pearson:2022:MappingDarkMatter,Walder+:2024:ProbingDarkMatter,Nibauer+Pearson:2026:TestingDarkMatter,Wu+:2026:PotamidesMappingDark,Chemaly+:2026:ConstraintsPopulationLevel}. %
        \citet{Starkman+:2026:EuclidQ1} extend these searches to \euclid{} imaging, while \rubinlsst{}, \gls{ARRAKIHS}, and \gls{Roman} will find many more streams \citep{Ivezic+:2019:LSSTScienceDrivers,Guzman+:2022:ARRAKIHS,Spergel+:2015:WideFieldInfrarRedSurvey}. %
        Our test applies to globular-cluster and dwarf-galaxy streams, the latter of which are easier to detect in extragalactic imaging. %

    % sec:discussion:outlook (end)

% sec:discussion (end)

\section{Conclusions}\label{sec:conclusions}

    In this paper we develop a novel framework for using observations of convexities in tracks of stellar streams to constrain the geometry of dark matter halos and distinguish the particle nature of dark matter. %
    We then derive the geometric constraints at the orbit level (\autoref{sec:diagnostic}), and carry them to streams via a width-bound geometric argument (\autoref{sec:streams:match_orbit}). %
    In \autoref{sec:cosmology} we build a population-level binomial test of \gls{CDM} against \gls{SIDM} and its cross-section by comparing the observed convexity rate $\lamObs$ to the predicted \gls{CDM} and \gls{SIDM} rates $\lamC, \lamSI$. %
    A catalog of convexities in extragalactic streams, like those current large-scale surveys will build, can favor \gls{CDM} or \gls{SIDM} by at least $2\sigma$, more likely $3$--$5\sigma$. %
    Of the stream-based methods, ours assumes the least: no parametric potential, progenitor model, or halo-shape prior. %

    We summarize the main results below. %

    \begin{itemize}[leftmargin=5pt, labelsep=0.5em]
        \item \emph{Geometric falsification.} %
        Detections of stream convexity falsifies any \gls{LoS}-axisymmetric, centrally-concentrated potential. %
        The result holds at the orbit level (\autoref{sec:diagnostic}), for thin streams, and for dwarf-galaxy progenitors via a width-bound geometric argument (\autoref{sec:streams:match_orbit}).  %
        \item \emph{Binomial cosmology test.} %
        Aggregating per-stream measurements across a survey of streams gives a binomial likelihood for \gls{CDM} vs \gls{SIDM} halo-shape distributions (\autoref{sec:cosmology}). %
        $N\!\sim\!\input{output/cosmology/n_required_fiducial_range.tex}$ extragalactic streams suffice for $2\sigma$--$5\sigma$ at challenging observational rates $(\lamC, \eta=\lamSI/\lamC)\!\sim\!(\lamCval,\input{output/cosmology/eta_fid.tex})$, and possibly as low as $\sim\!\input{output/cosmology/n_required_optimistic_range.tex}$ with more favorable detection rates $(\input{output/cosmology/lambda_det_cdm.tex}\%, \input{output/cosmology/eta_det.tex})$. These are within reach of catalogs from \euclid, \textit{Rubin}, and \textit{Roman}. %
        \item \emph{A cross-section constraint.} %
        The rate ratio $\eta=\lamSI/\lamC$ is sensitive to \gls{SIDM} cross-sections of roughly $\sigma/m\gtrsim\qty{1}{\square\centi\meter\per\gram}$. %
        Sensitivity below $\sim\qty{1}{\square\centi\meter\per\gram}$ is contingent on \gls{CDM} and \gls{SIDM} halos retaining different shapes in elliptical hosts despite the effects of baryons (\autoref{sec:cosmology:microphysics}). %
        \item \emph{Shape-free population test.} %
        The cosmology test needs minimal priors: none about the progenitor and no parametric potential. %
        The data can test any predicted halo-shape distribution. %
        The price of assuming so little is low per-stream signal, so the method needs a large stream catalog. %
        \item \emph{Complementary to higher-signal methods.} %
        Our binary check is parameter-free, and any subsequent potential fit by any other method must be consistent with it. %
        \item \emph{Upcoming survey data.} %
        The diagnostic uses only the projected track and the projected center of mass. %
        It therefore applies directly to the arriving extragalactic stream catalogs from \euclid, \gls{Rubin}, and \gls{Roman}. %
    \end{itemize}

    Even without detailed modeling of individual streams or their host galaxy potentials, stellar streams in aggregate are a powerful probe of the nature of dark matter. %

% sec:conclusions (end)

\newpage
%% Acknowledgments
\begin{acknowledgments}
    % Alphabetical
    % Lina Necib
    L.N. received support from NSF, via CAREER award AST-2337864. L.N. is also supported by the Sloan Fellowship. 
    $\cdot$
    % Jake Nibauer
    J.N. was supported in part by a National Science Foundation Graduate Research Fellowship, Grant No. DGE 2039656. Any opinions, findings, and conclusions or recommendations expressed in this material are those of the author(s) and do not necessarily reflect the views of the National Science Foundation. %
    $\cdot$
    % Sarah Pearson (& Sirui Wu, under the same grant)
    This work was supported by a research grant (VIL53081) from VILLUM FONDEN. This work was also co-funded by the European Union (ERC, BeyondSTREAMS, 101115754) grant. Views and opinions expressed are however those of the author(s) only and do not necessarily reflect those of the European Union or the European Research Council. Neither the European Union nor the granting authority can be held responsible for them.
    $\cdot$
    % Nathaniel Starkman
    Support for this work was provided by The Brinson Foundation through a Brinson Prize Fellowship grant to author N. S. 

    % AI Assistance Disclosure
    \vspace{10pt}
    This work made use of Anthropic's Claude (Sonnet and Opus models) (\url{https://www.anthropic.com/claude}) as an assistive tool for code development, manuscript editing, suggesting relevant literature, and discussion of conceptual aspects of the analysis. All suggested references were checked against the original sources by the authors, and all scientific content, analysis choices, and conclusions are the authors' own.

\end{acknowledgments}

\software{
    \texttt{numpy} \citep{numpy},
    \texttt{astropy} \citep{astropy},
    \texttt{matplotlib} \citep{matplotlib},
    \texttt{jax} \citep{jax},
    \texttt{unxt} \citep{unxt},
    \texttt{coordinax} \citep{coordinax},
    \texttt{galax} \citep{galax},
    \texttt{phasecurvefit} \citep{Starkman+2026:phasecurvefit}.
}

\section*{Author Contributions}\label{sec:author_contributions}
    (Alphabetical)

    L.~Necib: Writing -- review \& editing.

    J.~Nibauer: Formal analysis, Investigation, Writing -- review \& editing.

    S.~Pearson: Writing -- review \& editing.

    N.~Starkman: Conceptualization, Methodology, Formal analysis, Software, Investigation, Visualization, Writing -- original draft, Writing -- review \& editing.

    S.~Wu: Conceptualization, Investigation, Software (supporting), Visualization, Writing -- review \& editing.

% sec:author_contributions (end)

%% Bibliography
\bibliography{main}
\bibliographystyle{aasjournalv7}

\appendix

\section{Robustness of the Ridge-Line Diagnostic}\label{sec:ridgeline_robustness}

    This appendix further develops three aspects of the robustness of the ridge-line diagnostic introduced in \autoref{sec:streams:match_orbit}. %
    \autoref{sec:ridgeline_robustness:width_bound} derives the width bound of \cref{eq:streams:width_bound} everywhere in the stream, not only at its edges. %
    \autoref{sec:ridgeline_robustness:smoothing} shows that $\Astr$ is not particularly sensitive to how we estimate the ridge line from the imaging. %
    \autoref{sec:triaxial:streakline} confirms that the orbit-level results of \autoref{sec:triaxial_potentials:orbits} also apply to the observed stream track. %
    
    \subsection{The Width Bound for Every Track}\label{sec:ridgeline_robustness:width_bound}

        \Cref{eq:streams:width_bound} bounds the ridge-line--orbit tangent mismatch by a width gradient. %
        In \autoref{sec:streams:match_orbit} this is done at the stream's edge, but the more general result applies to every track within the stream. %

        Let a track neighboring the ridge line track have perpendicular offset $\xi(\ssky)$, with $|\xi|\le W$, so that ${\xsky}_2 = \xsky + \xi\,\nphat$. %
        Differentiating and using \cref{eq:theory:path:curvature_normal_2d} gives %
        \begin{equation}\label{eq:streams:neighbor_tangent}
            \deriv[\ssky]{{\xsky}_2}
            = \bigl(1-\curvaturesky\,\xi\bigr)\,\tphat
            + \bigl(\pderiv**[\ssky]{\xi}\bigr)\,\nphat , %
        \end{equation}
        so the neighboring track tilts from the ridge line by an angle $\delta\theta_{12}$, %
        \begin{equation}\label{eq:streams:neighbor_angle}
            \tan\delta\theta_{12}
            = \frac{\pderiv**[\ssky]{\xi}}{1-\curvaturesky\,\xi}
            \simeq \pderiv**[\ssky]{\xi} , %
        \end{equation}
        where the approximation holds when the stream width is small compared with the local radius of curvature, $\curvaturesky W \ll 1$. %
        Setting $\xi = \pm W$ gives the result from \autoref{sec:streams:match_orbit}, $\tan\delta\theta_{12} = \pm\,\pderiv**[\ssky]{W}$. %
        Away from major features such as epicyclic folds, the tracks within a stream do not cross, so each track remains at an approximately fixed fraction of the stream width: $\xi = fW$, with $|f| \le 1$ and $f$ varying slowly along the stream. %
        Thus, $\pderiv**[\ssky]{\xi} \simeq f\,\pderiv**[\ssky]{W}$, and the tilt of every track relative to the ridge line is bounded by $|\pderiv**[\ssky]{W}|$, recovering \cref{eq:streams:width_bound}. %

    % sec:ridgeline_robustness:width_bound (end)

    \subsection{Maintaining Curvature Alignment from Physics to Observations}\label{sec:ridgeline_robustness:smoothing}

        In this section we show that the difference between the physical mechanisms that produce smooth streams and the observational and numerical means by which we fit a smooth ridge line does not change the diagnostic $\Astr$. %
        Considering the spatial aspect of Lagrange-point release, a tidal stream should not be smooth: epicyclic loops and overdensities of wavenumber $k_{\mrm{epi}}$ should structure it about its ridge line \citep{Kupper+:2010:TidalTailsStar,Kupper+:2012:MoreStructureTidal}. %
        However, the spread in release velocities at the Lagrange-point release mixes the epicyclic phases and damps the loops \citep{Bovy:2014:DynamicalModelingTidal} sufficiently that the concavity changes of the loops are not observed. %
        When modeling the stream track from observations, a fit over an along-stream window $\Delta \ssky$ suppresses the loops by $\operatorname{sinc}(k_{\mrm{epi}}\Delta \ssky/2)$, so any window with $\Delta \ssky \gtrsim 2\pi/k_{\mrm{epi}}$ removes them. %
        These two different mechanisms both suppress small-scale changes in the concavity without impacting any larger-scale changes. %

        This ridge line modeled from observations will differ from the ``true'' ridge line, and we want to understand under what conditions their concavities agree. %
        Write the difference between them as a shift $\delta_{\mrm{sm}}\xsky$. %
        Let $\mbf{n}_0$ be the ridge-line tangent rotated by $\qty[retain-explicit-plus]{+90}{\degree}$. %
        Unlike $\nphat$, which reverses at an inflection point, $\mbf{n}_0$ varies more smoothly along the track. %
        In terms of $\mbf{n}_0$ the alignment \cref{eq:theory:path:alignment_def} is %
        \begin{equation}\label{eq:robust:smoothing:alignment_sign}
            \Astr(\ssky)
            =
            -\operatorname{sgn}\!\left[
                \bigl((\xsky-\xcom)\cdot\mbf{n}_0\bigr)\,
                \bigl(\pderiv**[\ssky]<2>{\xsky}\cdot\mbf{n}_0\bigr)
            \right], %
        \end{equation}
        where the first factor is the signed perpendicular distance from the \gls{CoM} to the tangent line and the second is the signed curvature. %
        The sign is preserved when $\delta_{\mrm{sm}}\xsky$ and $\pderiv**[\ssky]<2>{\delta_{\mrm{sm}}\xsky}$ are small compared with the two factors. %
        A shift of order the stream width does not change either, so $\Astr$ is the same for both. %

        The exceptions are when one of the factors vanishes. %
        \autoref{sec:streams:match_orbit} already shows these cases are identifiable in imaging, so we exclude them.
        At an inflection point the curvature vanishes and $\Astr$ will change sign; the exact location of the change depends on the ridge line model and does not matter in this work where we only care only that $\Astr$ changes. %
        For radial infall the perpendicular distance vanishes (\cref{eq:streams:match_orbit:sign_robust}), but this configuration does not form streams \citep{Hendel:2015:TidalDebrisMorphology}. %
        For a line of sight near the orbital plane the projected stream can satisfy these cases, which is why the width conditions of \autoref{sec:streams:match_orbit} are important. %

    % sec:ridgeline_robustness:smoothing (end)

    \subsection{Streak-lines in Triaxial Potentials}\label{sec:triaxial:streakline}

        % Transfer in triaxial hosts
        Orbits in triaxial potentials are not uniformly concave, and \cref{eq:theory:orbit:triaxial:alignment} gives $\mathcal{A}_{\mrm{orb}}=\pm1$ along the track (\autoref{sec:triaxial_potentials:orbits}). %
        We show here that the ridge-line diagnostic $\mathcal{A}_{\mrm{str}}(\ssky)$ reproduces that pattern pointwise, so the orbit-level results of \autoref{sec:triaxial_potentials:orbits} carry over to observed convex segments. %

        % Sign stability
        The perpendicular-distance argument applies unchanged, with the orbit-level prediction $\mathcal{A}_{\mrm{orb}}(\ssky)$ of \cref{eq:theory:orbit:triaxial:alignment} in place of the constant $+1$. %
        We evaluate it at the position, \gls{LoS} coordinate, and velocity direction of the star on the ridge line at $(t_0,\tau(\ssky))$. %
        Non-axisymmetric forces enter the linearized flow at the same order as axisymmetric ones, so the cold-release scalings $W\sim\delta r$ and $|\pderiv**[\ssky]{W}|\sim\delta r/r$ carry over. %
        The ridge-line and orbit alignments therefore agree, $\mathcal{A}_{\mrm{str}}(\ssky) = \mathcal{A}_{\mrm{orb}}(\ssky)$, wherever $\|(\xsky^{\pm})_{\perp}\| \gg \delta r(\tau(\ssky))$. %
        Morphological selection \citep{Hendel:2015:TidalDebrisMorphology} excludes the near-radial regime as before, so thin streams satisfy this on most of their length. %
        Outside the near-tangent strip the results of \autoref{sec:triaxial_potentials:orbits} apply to the ridge line, and orbit and ridge statements are interchangeable to $\mathcal{O}(\delta r/r)$. %

    % sec:triaxial:streakline (end)

% sec:ridgeline_robustness (end)

\section{Triaxiality in the Density vs Potential} \label{sec:triaxiality_density_vs_potential}

    We work with the potential form $\Phi(m)$ rather than a self-consistent triaxial density $\rho(\xvec) = \rho_0(m)$. %
    For a triaxial density, the potential is typically computed via the Chandrasekhar ellipsoidal integral formalism \citep{Chandrasekhar:1969:EllipsoidalFiguresEquilibrium, Binney+Tremaine:2008:GalacticDynamics}. %
    The result has neither the form $\Phi(m)$ nor the closed-form force law of \cref{eq:theory:orbit:triaxial:g}, so the analytic treatment of \autoref{sec:triaxial_potentials:orbits} does not apply directly. %
    The gravitational potential is nonetheless always rounder than the density sourcing it \citep{Binney+Tremaine:2008:GalacticDynamics}, so a given density triaxiality produces a weaker triaxiality in the potential. %
    In simulations for \autoref{sec:cosmology}, we instead integrate orbits in the full self-consistent triaxial potential computed from the triaxial \gls{NFW} density via the Chandrasekhar formalism, with the density axis ratios $(b/a,\,c/a)$ motivated by cosmological $N$-body simulations. %

% sec:triaxiality_density_vs_potential (end)

\section{The Convexity Rate Grid in Full} \label{sec:convexity_rate_grid_full}

    \begin{figure*}[t]
        \centering
        \includegraphics[width=\linewidth]{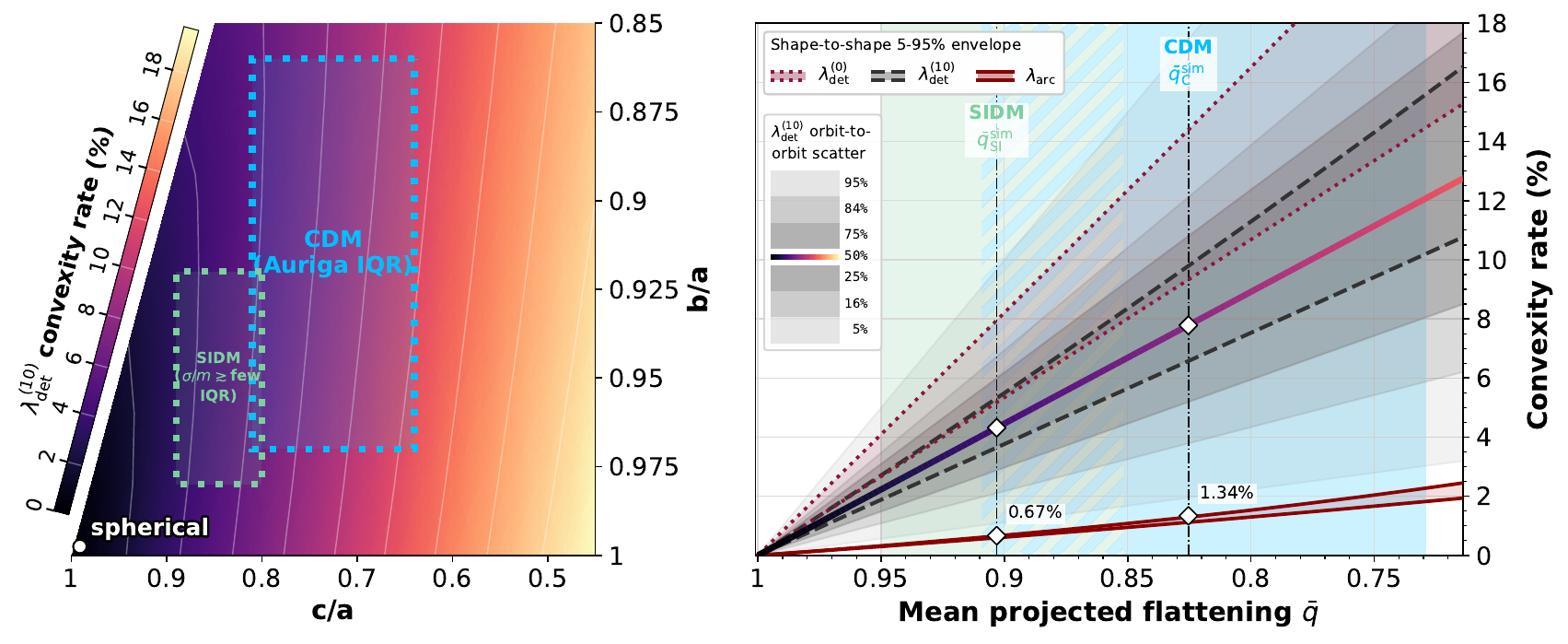}
        \caption{%
            Full version of \autoref{fig:cosmology:convexity_rate_vs_flattening}. %
            Burgundy dotted curves add $\lamdadet{0}$, the ideal rate if every convex segment is recovered. %
            Nested gray bands show additional orbit-to-orbit ranges. %
            Blue and green vertical bands show the central $\input{output/cosmology/band_pctl.tex}\%$ ranges of projected flattening predicted for the \gls{CDM} and \gls{SIDM} populations; hatching marks their overlap. %
            \label{fig:cosmology:convexity_rate_grid_full}
        }
    \end{figure*}

    For readability, \autoref{fig:cosmology:convexity_rate_vs_flattening} omits some of the rate and population-distribution detail. %
    This full version adds $\lamdadet{0}$, the ideal rate obtained when every convex segment is recovered; the outer orbit-to-orbit quantile bands; and blue and green vertical bands showing the central $\input{output/cosmology/band_pctl.tex}\%$ ranges of projected flattening for the \gls{CDM} and \gls{SIDM} populations. %
    Hatching marks where the two projected-flattening ranges overlap. %

    Because the \gls{SIDM} population lies close to the spherical limit, its predicted rate depends steeply on $\bar{q}$. %
    Shifting $\bar{q}_{\mrm{SI}}^{\rm sim}$ by $\input{output/cosmology/qbar_shift.tex}$, well within the variation between different \gls{SIDM} simulations, changes $\lamSI$ by $\sim\!\input{output/cosmology/lambda_sidm_shift.tex}\%$. %

% sec:convexity_rate_grid_full (end)

\end{document}

%% file: output/cosmology/lambda_arc_cdm.tex
% GENERATED by src/scripts/make_variables.py from variables.yml. DO NOT EDIT.
1.3%

%% file: output/cosmology/lambda_arc_sidm.tex
% GENERATED by src/scripts/make_variables.py from variables.yml. DO NOT EDIT.
0.67%

%% file: output/gd1/n_views.tex
% GENERATED by src/scripts/make_variables.py from variables.yml. DO NOT EDIT.
\num{e4}%

%% file: output/gd1/frac_any_convex.tex
% GENERATED by src/scripts/make_variables.py from variables.yml. DO NOT EDIT.
23.1%

%% file: output/gd1/frac_det_ell_min.tex
% GENERATED by src/scripts/make_variables.py from variables.yml. DO NOT EDIT.
3.2%

%% file: output/cosmology/ell_min.tex
% GENERATED by src/scripts/make_variables.py from variables.yml. DO NOT EDIT.
\qty{10}{\kilo\parsec}%

%% file: output/gd1/frac_fully_convex.tex
% GENERATED by src/scripts/make_variables.py from variables.yml. DO NOT EDIT.
1.6%

%% file: output/gd1/theta.tex
% GENERATED by src/scripts/make_variables.py from variables.yml. DO NOT EDIT.
\qty{113.2}{\degree}%

%% file: output/gd1/phi.tex
% GENERATED by src/scripts/make_variables.py from variables.yml. DO NOT EDIT.
\qty{19}{\degree}%

%% file: output/gd1/convex_frac.tex
% GENERATED by src/scripts/make_variables.py from variables.yml. DO NOT EDIT.
100%

%% file: output/triaxial/t_integrate.tex
% GENERATED by src/scripts/make_variables.py from variables.yml. DO NOT EDIT.
\qty{5}{\giga\year}%

%% file: output/triaxial/m_tot.tex
% GENERATED by src/scripts/make_variables.py from variables.yml. DO NOT EDIT.
\qty{e12}{\Msun}%

%% file: output/triaxial/r_s.tex
% GENERATED by src/scripts/make_variables.py from variables.yml. DO NOT EDIT.
\qty{14}{\kilo\parsec}%

%% file: output/triaxial/q1.tex
% GENERATED by src/scripts/make_variables.py from variables.yml. DO NOT EDIT.
0.85%

%% file: output/triaxial/q2.tex
% GENERATED by src/scripts/make_variables.py from variables.yml. DO NOT EDIT.
0.7%

%% file: output/triaxial/orbit1_ecc.tex
% GENERATED by src/scripts/make_variables.py from variables.yml. DO NOT EDIT.
[0.06, 0.08]%

%% file: output/triaxial/orbit2_ecc.tex
% GENERATED by src/scripts/make_variables.py from variables.yml. DO NOT EDIT.
[0.48, 0.57]%

%% file: output/triaxial/r_near.tex
% GENERATED by src/scripts/make_variables.py from variables.yml. DO NOT EDIT.
\qty{25}{\kilo\parsec}%

%% file: output/triaxial/orbit1_peak_convex.tex
% GENERATED by src/scripts/make_variables.py from variables.yml. DO NOT EDIT.
55%

%% file: output/triaxial/orbit2_peak_convex.tex
% GENERATED by src/scripts/make_variables.py from variables.yml. DO NOT EDIT.
28%

%% file: output/cosmology/n_orbits.tex
% GENERATED by src/scripts/make_variables.py from variables.yml. DO NOT EDIT.
\num{5000}%

%% file: output/cosmology/t_integrate.tex
% GENERATED by src/scripts/make_variables.py from variables.yml. DO NOT EDIT.
\qty{5}{\giga\year}%

%% file: output/cosmology/r_range.tex
% GENERATED by src/scripts/make_variables.py from variables.yml. DO NOT EDIT.
\qtyrange{35}{100}{\kilo\parsec}%

%% file: output/cosmology/e_range.tex
% GENERATED by src/scripts/make_variables.py from variables.yml. DO NOT EDIT.
[0, 0.75]%

%% file: output/cosmology/i_range.tex
% GENERATED by src/scripts/make_variables.py from variables.yml. DO NOT EDIT.
\qtyrange{0}{360}{\degree}%

%% file: output/cosmology/n_views.tex
% GENERATED by src/scripts/make_variables.py from variables.yml. DO NOT EDIT.
\num{7200}%

%% file: output/cosmology/r_exclude.tex
% GENERATED by src/scripts/make_variables.py from variables.yml. DO NOT EDIT.
\qty{10}{\kilo\parsec}%

%% file: output/cosmology/qbar_turnover_ba_range.tex
% GENERATED by src/scripts/make_variables.py from variables.yml. DO NOT EDIT.
\numrange{0.90}{0.94}%

%% file: output/cosmology/qbar_turnover_ca.tex
% GENERATED by src/scripts/make_variables.py from variables.yml. DO NOT EDIT.
0.81%

%% file: output/cosmology/qbar_turnover_qbar_peak.tex
% GENERATED by src/scripts/make_variables.py from variables.yml. DO NOT EDIT.
0.884%

%% file: output/cosmology/qbar_turnover_ba.tex
% GENERATED by src/scripts/make_variables.py from variables.yml. DO NOT EDIT.
0.92%

%% file: output/cosmology/qbar_turnover_qbar_oblate.tex
% GENERATED by src/scripts/make_variables.py from variables.yml. DO NOT EDIT.
0.876%

%% file: output/cosmology/qbar_turnover_rate_peak.tex
% GENERATED by src/scripts/make_variables.py from variables.yml. DO NOT EDIT.
5.8%

%% file: output/cosmology/qbar_turnover_rate_oblate.tex
% GENERATED by src/scripts/make_variables.py from variables.yml. DO NOT EDIT.
6.8%

%% file: output/cosmology/band_pctl.tex
% GENERATED by src/scripts/make_variables.py from variables.yml. DO NOT EDIT.
5\text{--}95%

%% file: output/cosmology/eta_region_mean.tex
% GENERATED by src/scripts/make_variables.py from variables.yml. DO NOT EDIT.
0.57%

%% file: output/cosmology/n_required_ratio.tex
% GENERATED by src/scripts/make_variables.py from variables.yml. DO NOT EDIT.
1.42%

%% file: output/cosmology/window.tex
% GENERATED by src/scripts/make_variables.py from variables.yml. DO NOT EDIT.
\qty{2}{\giga\year}%

%% file: output/cosmology/cdm_ba_range.tex
% GENERATED by src/scripts/make_variables.py from variables.yml. DO NOT EDIT.
\numrange{0.86}{0.97}%

%% file: output/cosmology/cdm_ca_range.tex
% GENERATED by src/scripts/make_variables.py from variables.yml. DO NOT EDIT.
\numrange{0.64}{0.81}%

%% file: output/cosmology/auriga_ca_median.tex
% GENERATED by src/scripts/make_variables.py from variables.yml. DO NOT EDIT.
0.73%

%% file: output/cosmology/sidm_relaxation.tex
% GENERATED by src/scripts/make_variables.py from variables.yml. DO NOT EDIT.
0.441%

%% file: output/cosmology/sidm_ca_offset.tex
% GENERATED by src/scripts/make_variables.py from variables.yml. DO NOT EDIT.
16.3%

%% file: output/cosmology/sidm_ba_range.tex
% GENERATED by src/scripts/make_variables.py from variables.yml. DO NOT EDIT.
\numrange{0.92}{0.98}%

%% file: output/cosmology/sidm_ca_range.tex
% GENERATED by src/scripts/make_variables.py from variables.yml. DO NOT EDIT.
\numrange{0.80}{0.89}%

%% file: output/cosmology/lambda_det_cdm.tex
% GENERATED by src/scripts/make_variables.py from variables.yml. DO NOT EDIT.
7.8%

%% file: output/cosmology/lambda_det_sidm.tex
% GENERATED by src/scripts/make_variables.py from variables.yml. DO NOT EDIT.
4.3%

%% file: output/cosmology/eta_fid.tex
% GENERATED by src/scripts/make_variables.py from variables.yml. DO NOT EDIT.
0.50%

%% file: output/cosmology/wald_error_1sigma.tex
% GENERATED by src/scripts/make_variables.py from variables.yml. DO NOT EDIT.
20%

%% file: output/cosmology/wald_error_3sigma.tex
% GENERATED by src/scripts/make_variables.py from variables.yml. DO NOT EDIT.
3.1%

%% file: output/cosmology/worked_n.tex
% GENERATED by src/scripts/make_variables.py from variables.yml. DO NOT EDIT.
3.5\times10^{3}%

%% file: output/cosmology/worked_k_cdm.tex
% GENERATED by src/scripts/make_variables.py from variables.yml. DO NOT EDIT.
47%

%% file: output/cosmology/worked_p_sidm.tex
% GENERATED by src/scripts/make_variables.py from variables.yml. DO NOT EDIT.
\num{9.5e-06}%

%% file: output/cosmology/worked_z_sidm.tex
% GENERATED by src/scripts/make_variables.py from variables.yml. DO NOT EDIT.
4.3%

%% file: output/cosmology/worked_k_sidm.tex
% GENERATED by src/scripts/make_variables.py from variables.yml. DO NOT EDIT.
23%

%% file: output/cosmology/worked_p_cdm.tex
% GENERATED by src/scripts/make_variables.py from variables.yml. DO NOT EDIT.
\num{8.2e-05}%

%% file: output/cosmology/worked_z_cdm.tex
% GENERATED by src/scripts/make_variables.py from variables.yml. DO NOT EDIT.
3.8%

%% file: output/cosmology/worked_z_sidm_wald.tex
% GENERATED by src/scripts/make_variables.py from variables.yml. DO NOT EDIT.
4.9%

%% file: output/cosmology/worked_z_cdm_wald.tex
% GENERATED by src/scripts/make_variables.py from variables.yml. DO NOT EDIT.
3.5%

%% file: output/cosmology/worked_n_sigma.tex
% GENERATED by src/scripts/make_variables.py from variables.yml. DO NOT EDIT.
2.0%

%% file: output/cosmology/catalog_sizes.tex
% GENERATED by src/scripts/make_variables.py from variables.yml. DO NOT EDIT.
10^{2},10^{3},10^{4}%

%% file: output/cosmology/n_euclid_min.tex
% GENERATED by src/scripts/make_variables.py from variables.yml. DO NOT EDIT.
\num{2300}%

%% file: output/cosmology/n_required_fiducial.tex
% GENERATED by src/scripts/make_variables.py from variables.yml. DO NOT EDIT.
\num{850}, \num{3400}, \num{7700}, \num{21000}%

%% file: output/cosmology/n_required_optimistic.tex
% GENERATED by src/scripts/make_variables.py from variables.yml. DO NOT EDIT.
\num{180}, \num{740}, \num{1700}, \num{4600}%

%% file: output/cosmology/eta_det.tex
% GENERATED by src/scripts/make_variables.py from variables.yml. DO NOT EDIT.
0.55%

%% file: output/cosmology/lambda_det_over_fiducial.tex
% GENERATED by src/scripts/make_variables.py from variables.yml. DO NOT EDIT.
5.8%

%% file: output/cosmology/n_required_optimistic_3sigma.tex
% GENERATED by src/scripts/make_variables.py from variables.yml. DO NOT EDIT.
\num{1700}%

%% file: output/cosmology/eta_lo.tex
% GENERATED by src/scripts/make_variables.py from variables.yml. DO NOT EDIT.
0.10%

%% file: output/cosmology/eta_hi.tex
% GENERATED by src/scripts/make_variables.py from variables.yml. DO NOT EDIT.
0.70%

%% file: output/cosmology/n_required_3sigma_eta_lo.tex
% GENERATED by src/scripts/make_variables.py from variables.yml. DO NOT EDIT.
\num{1400}%

%% file: output/cosmology/n_required_3sigma_eta_hi.tex
% GENERATED by src/scripts/make_variables.py from variables.yml. DO NOT EDIT.
\num{25000}%

%% file: output/cosmology/n_sigma_local_group.tex
% GENERATED by src/scripts/make_variables.py from variables.yml. DO NOT EDIT.
0.34%

%% file: output/cosmology/eta_sidm_regime.tex
% GENERATED by src/scripts/make_variables.py from variables.yml. DO NOT EDIT.
0.5%

%% file: output/cosmology/n_required_fiducial_range.tex
% GENERATED by src/scripts/make_variables.py from variables.yml. DO NOT EDIT.
\numrange{3400}{21000}%

%% file: output/cosmology/n_required_optimistic_range.tex
% GENERATED by src/scripts/make_variables.py from variables.yml. DO NOT EDIT.
\numrange{740}{4600}%

%% file: output/cosmology/qbar_shift.tex
% GENERATED by src/scripts/make_variables.py from variables.yml. DO NOT EDIT.
0.0031%

%% file: output/cosmology/lambda_sidm_shift.tex
% GENERATED by src/scripts/make_variables.py from variables.yml. DO NOT EDIT.
4%